\documentclass[prd,aps,amsfonts,eqsecnum,superscriptaddress,nofootinbib,longbibliography,notitlepage]{revtex4-1}

\usepackage{graphicx}
\usepackage{xcolor}
\usepackage{rotating}
\usepackage{amsmath,amssymb,graphics,amsthm,isomath}
\usepackage{bbm}

\usepackage[colorlinks=true, urlcolor=violet, linkcolor=blue, citecolor=red, hyperindex=true, linktocpage=true]{hyperref}
\usepackage[capitalise,compress]{cleveref}

\allowdisplaybreaks

\numberwithin{thm}{section}

\makeatletter
\renewcommand{\p@subsection}{}
\renewcommand{\p@subsubsection}{}
\makeatother

\usepackage{xcolor}
\usepackage{mathtools}

\def\ra{\rangle}
\def\la{\langle}

\def\bea{\begin{eqnarray}}
\def\eea{\end{eqnarray}}
\def\be{\begin{equation}}
\def\ee{\end{equation}}
\def\bes{\begin{subequations}}
\def\ees{\end{subequations}}
\def\bed{\begin{displaymath}}
\def\eed{\end{displaymath}}
\def\beal{\begin{aligned}}
\def\eeal{\end{aligned}}
\def\bew{\begin{widetext}}
\def\eew{\end{widetext}}
\def\beit{\begin{itemize}}
\def\eeit{\end{itemize}}
\def\bea{\begin{array}}
\def\eea{\end{array}}
\def\been{\begin{enumerate}}
\def\eeen{\end{enumerate}}

\newcommand{\eqnref}[1]{\eqref{#1}}
\newcommand{\figref}[1]{Figure\;\ref{#1}}
\newcommand{\tabref}[1]{Table\;\ref{#1}}
\newcommand{\secref}[1]{Section\;\ref{#1}}
\newcommand{\appref}[1]{Appendix\;\ref{#1}}

\def\o{\omega_c}

\def\th{\theta}\def\d{\delta}
\def\pa{\partial}
\def\ep{\epsilon}
\def\a{\alpha}

\def\Ge{\Gamma_{\text{e}}}
\def\Gph{\Gamma_{\text{ph}}}
\def\ga{\gamma}
\def\bw{\bar{w}}
\def\ba{\bar{a}}

\def\e{\text{e}}
\def\ph{\text{ph}}
\def\inc{\text{inc}}

\def\L{\Lambda}
\def\b{\boldsymbol}

\makeatletter
\DeclareRobustCommand{\cev}[1]{%
  \mathpalette\do@cev{#1}%
}
\newcommand{\do@cev}[2]{%
  \fix@cev{#1}{+}%
  \reflectbox{$\m@th#1\vec{\reflectbox{$\fix@cev{#1}{-}\m@th#1#2\fix@cev{#1}{+}$}}$}%
  \fix@cev{#1}{-}%
}
\newcommand{\fix@cev}[2]{%
  \ifx#1\displaystyle
    \mkern#23mu
  \else
    \ifx#1\textstyle
      \mkern#23mu
    \else
      \ifx#1\scriptstyle
        \mkern#22mu
      \else
        \mkern#22mu
      \fi
    \fi
  \fi
}

\makeatother

\newcommand{\ud}{\mathrm{d}}
\def\i{\text{i}}

\def\mC{\mathcal{C}}

\def\bbP{\mathbb{P}}

\def\mO{\mathcal{O}}

\newcommand{\norm}[1]{\left\lVert#1\right\rVert}

\usepackage{dsfont}

\begin{document}

\title{Electron-phonon hydrodynamics}

\author{Xiaoyang Huang}
\email{xiaoyang.huang@colorado.edu}
\affiliation{Department of Physics and Center for Theory of Quantum Matter, University of Colorado, Boulder CO 80309, USA}

\author{Andrew Lucas}
\email{andrew.j.lucas@colorado.edu}
\affiliation{Department of Physics and Center for Theory of Quantum Matter, University of Colorado, Boulder CO 80309, USA}

\begin{abstract}
We develop the theory of hydrodynamics of an isotropic Fermi liquid of electrons coupled to isotropic acoustic phonons, assuming that umklapp processes may be neglected.  At low temperatures, the fluid is approximately Galilean invariant; at high temperatures, the fluid is nearly relativistic; at intermediate temperatures, there are seven additional temperature regimes with unconventional thermodynamic properties and hydrodynamic transport coefficients in a three-dimensional system.  We predict qualitative signatures of electron-phonon fluids in incoherent transport coefficients, shear and Hall viscosity, and plasmon dispersion relations.  Our theory may be relevant for numerous quantum materials where strong electron-phonon scattering has been proposed to underlie a hydrodynamic regime, including $\mathrm{WTe}_2$, $\mathrm{WP}_2$, and $\mathrm{PtSn}_4$.
\end{abstract}

\date{\today}

\maketitle
\tableofcontents

\section{Introduction}

The hydrodynamics of correlated electron liquids, theorized many decades ago \cite{gurzhi}, has recently become increasingly observable in experiments in a broad range of materials, including graphene \cite{Bandurin_2016,Crossno_2016,ghahari,Krishna_Kumar_2017,Gallagher158,Berdyugin_2019,sulpizio,jenkins2020imaging,Ku_2020} and GaAs \cite{de_Jong_1995,Gusev_2018}; see \cite{Lucas_2018} for a recent review, and \cite{Guo_2017,Levitov_2016,Torre_2015,alekseev,Andreev_2011,Apostolov_2014,Forcella_2014,Tomadin_2014} for recent theoretical developments.  More speculatively, proposed evidence for hydrodynamics has been put forth in $\mathrm{WTe}_2$ \cite{vool2020imaging}, $\mathrm{WP}_2 $ \cite{Gooth_2018,coulter,osterhoudt2020evidence,Jaoui2018}, $\mathrm{PtSn}_4$ \cite{fu2018thermoelectric} and $\mathrm{PdCoO}_2$ \cite{Moll_2016,caleb}. In these more complicated material systems, the electron's Fermi surface is highly anisotropic (and may be multiple-sheeted).  Moreover, in many of these materials, it is believed that electron-phonon scattering cannot be neglected, and might even be largely responsible for the experimental signatures of hydrodynamics.  Can exotic hydrodynamics arise in these material systems, with phenomenology beyond the canonical (Galilean-invariant) hydrodynamics?

To begin to address this question, we revisit the hydrodynamics of a coupled fluid of electrons and acoustic phonons, in more than one spatial dimension $d>1$.  This problem has arisen  in the older solid-state literature \cite{steinberg1958,gurzhi1972}, with a recent work \cite{levchenko2020transport} revisiting this theory in light of the more recent experimental developments cited above.  The focus of this earlier literature is largely on low temperature dynamics, with a notable exception of \cite{gurevich1967,nielsen1969} studying high temperature phonon-dominated hydrodynamics.  The purpose of this paper is to show that there are at least 7 distinct temperature regimes of coupled electron-phonon fluids, exhibiting qualitatively distinct behaviors, that can arise at temperatures below the Fermi temperature in $d=2$, and at least 9 different temperature regimes in $d>2$.  These distinct temperature regimes arise even with the simplest electron-phonon scattering integrals, and ignoring a multitude of possible microscopic effects such as optical phonon scattering, anisotropy or multi-sheeted Fermi surfaces. These distinct temperature regimes are distinguishable when the ratio of the phonon velocity and the Fermi velocity is parametrically small.  In practice, this ratio might typically be around 0.01 to 0.1, which is small enough to distinguish at least a few of the different fluids we will describe (though distinguishing all 9 regimes cleanly would be difficult in principle and in practice).  The previous literature \cite{steinberg1958,nielsen1969,gurzhi1972,levchenko2020transport} which focuses on electron-phonon coupled fluids (as far as we could tell) studies only one of these temperature regimes, that arises at the very lowest temperatures.  

However, we will argue, based on our theory and experimental data from the relevant materials, that this is not the range of temperatures most relevant for understanding signatures of hydrodynamics in experiments in many of the material systems listed above.  The main results of this paper are the unique signatures of the hybrid electron-phonon fluid as a function of temperature and/or magnetic field dependence of numerous coefficients, including shear and Hall viscosity, incoherent conductivity, and plasmon dispersion.  As we will discuss, these phenomena can be probed experimentally in the near future via numerous experimental techniques, including conductance measurements in narrow channels or constrictions \cite{Guo_2017,Krishna_Kumar_2017}, imaging methods \cite{sulpizio,Ku_2020,jenkins2020imaging}, or the properties of collective modes such as plasmons \cite{abbamonte19,husain2020coexisting}.   Our predictions are not unique to one class of material such as $\mathrm{WP}_2$;  instead, they follow only from the assumption that electron-phonon (rather than electron-electron) scattering is responsible for electron hydrodynamics.

\section{Summary of results}

We now broadly summarize our results.  We first introduce the model of interest.  We consider an isotropic and weakly interacting electron Fermi liquid, with an isotropic dispersion relation.  We suppose that the Fermi temperature is $T_{\mathrm{F}}$, the Fermi momentum is $p_{\mathrm{F}}$, and the Fermi velocity is $v_{\mathrm{F}}$. We assume the electronic density of states at the Fermi energy is given by  
\begin{equation} \label{eq:nuidentity}
    \nu = \frac{\pa n}{\pa \mu}=\frac{\Omega_{d-1}}{(2\pi)^d}dp^{d-1}_{\mathrm{F}}\frac{\pa p_{\mathrm{F}}}{\pa \mu}=d\frac{n}{v_{\mathrm{F}}p_{\mathrm{F}}} = \frac{dn}{mv_{\mathrm{F}}^2} \sim \frac{p_{\mathrm{F}}^{d-1}}{v_{\mathrm{F}}}.
\end{equation} where $\Omega_{d-1}$ is the area of the unit sphere in $d-1$ dimensions and $m$ is an effective quasiparticle mass. Working in units where $k_{\mathrm{B}}=1$, the single-particle fermion dispersion relation near the Fermi surface is \begin{equation}
    \epsilon(\mathbf{p}) = T_{\mathrm{F}} + v_{\mathrm{F}}(|\mathbf{p}| - p_{\mathrm{F}}) + \frac{\pa_pv_{\mathrm{F}}}{2} (|\mathbf{p}| - p_{\mathrm{F}})^2 + \cdots 
\end{equation}  Similarly, we consider isotropic acoustic phonons whose frequency is given by 
\begin{equation}
    \omega(\mathbf{p}) = v_{\mathrm{ph}}|\mathbf{p}|.
\end{equation}
We restrict our calculations to temperatures $T\ll T_{\mathrm{F}}$, and will generally suppress all subleading corrections in the small parameter $T/T_{\mathrm{F}}$ except where stated. In usual metals ($T_{\mathrm{F}}\sim10000$ K) the criterion is impossible to violate, but for low density systems it is possible, e.g. in graphene \cite{Crossno_2016,Gallagher158} and GaAs \cite{Gusev_2018,PhysRevB.77.235437,Bernardi5291}. The system we consider is $d$-dimensional with $d>1$.  Experimental systems exist with both $d=2$ (exactly, or approximately) and $d=3$. 

For phonon modes to be effectively two-dimensional in a layered heterostructure, it is sufficient to have the thickness of the material in the third dimension larger than $\hbar v_{\ph}/k_{\mathrm{B}} T$. Although in principle this seems easily acheived in GaAs, where \cite{Gusev_2018} $T\approx 1$ K and $v_{\mathrm{ph}}\sim 4\times 10^3$ m/s is appropriate for the hydrodynamic regime (the desired thickness is of order 30 nm), in practice the phonons may not be sharply bound to the GaAs layer in a heterostructure.  Alternatively, one may consider van der Waals heterostructures as a natural experimental playground for two-dimensional electron-phonon hydrodynamics.  A possible complication in these systems (especially in suspended monolayer graphene) will be the addition of low energy flexural phonons \cite{flex1, flex2}, which we have not included in our calculation.\footnote{We nevertheless anticipate that the hydrodynamics of electrons interacting with flexural phonons is, in many respects, rather similar to what is described in this work, albeit with numerous crossover temperatures modified, and temperature depenendences modified.}

Note that all of the materials $\mathrm{WP}_2$, $\mathrm{WTe}_2$, $\mathrm{PtSn}_4$ and $\mathrm{PdCoO}_2$ are anisotropic.  Moreover, analyses of models with electron-electron scattering \cite{caleb} demonstrate new phenomena arising from anisotropy.  Nevertheless, we focus on the simpler setting of isotropic liquids in this paper.   As we will see, this simpler setting is already extremely complex.  We anticipate that nearly all of the \emph{qualitative} features of our model will continue to hold in anisotropic models, with the likely exception of the low temperature scaling of incoherent conductivity \cite{caleb}.

As such, we press on with our minimal model.  There are three dimensionless numbers that will prove particularly important in our analysis. First, there is the small parameter \begin{equation}\label{eqn:ba}
    \ba\equiv \frac{a}{p_{\mathrm{F}}}\equiv \frac{\pi T}{\sqrt{3}p_{\mathrm{F}}v_{\mathrm{F}}} ,
\end{equation}
which represents the smallness of thermal fluctuations of the electrons about their Fermi surface -- specifically, the enhancement of the momentum carried by thermal fluctuations of electrons.  Secondly, we have the ratio
\be
\bw\equiv \frac{w}{p_{\mathrm{F}}\sqrt{\nu/2}}\equiv \frac{I(d)}{p_{\mathrm{F}}\sqrt{\nu/2}}\sqrt{\frac{T^{d+1}}{v_{\ph}^{d+2}}},
\label{eqn:bw}
\ee
where $I(d)$ is a dimensionless number defined in (\ref{eqn:ph-norm}).  $\bw$ represents the ratio of the momentum carried by phonons to momentum carried by electrons in equilibrium.   As $\bw$ becomes larger with increasing temperature, the fluid's properties qualitatively change.  $\bw \ll 1 $ and $\bw \gg 1$ are both limits which can be achieved in experimental devices, as we will soon see.  Finally, there is the small ratio \begin{equation}
    r = \frac{v_{\mathrm{ph}}}{v_{\mathrm{F}}} \label{eq:defr}
\end{equation}
characterizing the small ratio of quasiparticle velocities between phonons and electrons.   We assume that the dispersion relation is sufficiently simple so that $T_{\mathrm{F}} \sim p_{\mathrm{F}}v_{\mathrm{F}}$, in which case we also obtain the following useful scaling relation:  \begin{equation}
    \bw \sim  \left(\frac{T}{T_{\mathrm{F}}}\right)^{(d+1)/2} r^{-(d+2)/2} \sim \frac{\ba^{(d+1)/2}}{r^{(d+2)/2}}.
\end{equation}

\begin{table}[t]
\begin{tabular}{|c|c|c|c|c|c|c|c|c|c|}
\hline
$T$ & $<T_1$ & $T_1$ to $T_2$ & $T_2$ to $T_3$ & $T_3$ to $T_4$ & $T_4$ to $T_{\text{BG}}$ & $T_{\text{BG}}$ to $T_6$ & $T_6$ to $T_7$ & $T_7$ to $T_8$ & $T_8$ to $T_{\mathrm{F}}$  \\
\cline{1-10}
regime & I & II & III & IV & V & VI & VII & VIII & IX \\
\cline{1-10}
sound velocity & \multicolumn{4}{|c|}{$ v_{\mathrm{F}}/\sqrt{d}$} & \multicolumn{2}{|c|}{$ v_{\mathrm{F}}/\bw\sqrt{d}$} & \multicolumn{3}{|c|}{$ v_{\ph}/\sqrt{d}$}\\
\cline{1-10}
charge current & coherent ($|J_{\inc}\ra\propto \ba$)  & \multicolumn{3}{|c|}{coherent ($|J_{\inc}\ra\propto \bw$)} & \multicolumn{5}{|c|}{incoherent}\\
\cline{1-10}
heat current  & \multicolumn{7}{|c|}{incoherent}  & \multicolumn{2}{|c|}{coherent} \\
\cline{1-10}
energy current & \multicolumn{4}{|c|}{coherent} & \multicolumn{4}{|c|}{incoherent} & coherent\\
\cline{1-10}
electron scattering & \multicolumn{5}{|c|}{small angle} & \multicolumn{4}{|c|}{large angle}\\
\cline{1-10}
bulk viscosity & \multicolumn{2}{|c|}{by $|2\ra_{\e}$} & by $|1\ra_{\ph}$ &   \multicolumn{6}{|c|}{by $|1\ra_{\e}$} \\
\cline{1-10}
$\Sigma$ v.s. $\eta$ & \multicolumn{4}{|c|}{$\Sigma< \eta$} & \multicolumn{5}{|c|}{$\Sigma> \eta$} \\
\hline
\end{tabular}
\caption{A summary of the hydrodynamic properties of a coupled electron-phonon fluid.  The ratio $r$ is defined in (\ref{eq:defr}) and the temperature crossovers are defined in (\ref{eq:defT}).}
\label{tab:sum}
\end{table}

The main result of this paper is the calculation of hydrodynamic transport coefficients in electron-phonon hydrodynamics. We summarize the results in \tabref{tab:sum}.  As a function of temperature $T$, in $d>2$ spatial dimensions there is a zoo of temperature scales at which certain properties change, which we list below:
\begin{subequations}\label{eq:defT}\begin{align}
    T_1 &=r^{\frac{d+2}{d-1}}T_{\mathrm{F}}, \\
    T_2 &= r^{\frac{d-1}{d-2}}T_{\mathrm{F}}, \\
    T_3 &= r^{\frac{d}{d-1}}T_{\mathrm{F}}, \\
    T_4 &= r^{\frac{d+2}{d+1}}T_{\mathrm{F}}, \\
    T_5 &= T_{\text{BG}} = rT_{\mathrm{F}}, \\
    T_6 &=r^{\frac{d}{d+1}}T_{\mathrm{F}}, \\
    T_7 &= r^{\frac{d-2}{d-1}}T_{\mathrm{F}},\\
    T_8 &= r^{\frac{d-2}{d+1}}T_{\mathrm{F}}.
\end{align}\end{subequations} 
When temperature is low enough ($T<T_1$), the phonon modes essentially become irrelevant, except for providing a scattering mechanism for the electrons.  This is the regime studied in  previous papers \cite{steinberg1958,levchenko2020transport}.

However, above temperature $T_1$, the phonons begin to play a more interesting role.  For the discussion that follows, let us assume the ballpark estimates $T_{\mathrm{F}} \sim 10000$ K and $r \approx 0.03$, and consider a metal in $d=3$.  In this cartoon metal, $T_1\sim 2$ K.  When $T>T_1$, the phonons begin to dominate the incoherent part of the charge conductivity, which now becomes anomalously large, even for an isotropic fluid.\footnote{Note that in an anisotropic fluid \cite{caleb}, the incoherent conductivity will be large even in the lower temperature regimes, in which case this temperature scale becomes more dramatic.}  However, in an isotropic fluid, the effect will be fairly small.  At temperature $T_2$ (about 10 K), the bulk viscosity becomes dominated by phonons, although again the correction is quite small.  At $T_3$ (about 60 K) and beyond, the bulk viscosity becomes dominated by electrons again, but is not suppressed for any electronic dispersion relation.\footnote{If the electrons had $\epsilon\propto p^2$, then the bulk viscosity (in the absence of phonons carrying momentum) exactly vanishes.  For $T<T_1$ this implies the bulk viscosity is much smaller than it would be for a generic Fermi liquid.  Admittedly, this effect is very hard to see experimentally; this is simply an interesting theoretical observation.} 

The fluid completely begins to change its character at $T_4$ (in our cartoon, about 140 K).  At this temperature, $\bw \sim 1$, and so the momentum of the coupled fluid begins to become phonon-dominated.  This leads to a number of crucial changes, including a rapid \emph{decrease} in the sound speed ($v_{\mathrm{s}} \sim T^{-(d+1)/2}$), a large incoherent electrical conductivity, and a sound mode whose decay is dominated by incoherent conductivity rather than viscosity.  As we describe later, such changes may be visible in experiments by studying changes to the plasmon dispersion relation.  At $T_5=T_{\mathrm{BG}}$, the Bloch-Gr\"uneisen temperature,\footnote{The Bloch-Gr$\ddot{\text{u}}$neisen temperature $T_{\text{BG}}$ arises when the phonon-limited resistivity shows a dramatic change from $\rho\sim T^{d+2}$ to $\rho\sim T$ \cite{PhysRevLett.105.256805} (in the absence of phonon drag, at least). It can be understood from the bosonic nature of phonon modes: when $T<T_{\text{BG}}$, the volume of phonon momentum integration is restricted to the $T/v_{\ph}$ window; when $T>T_{\text{BG}}$, the phonons are more like a classical gas with equipartition distribution.} there are surprisingly few changes to the hydrodynamic properties of the fluid, beyond a possible sharp decrease in the shear viscosity.  In our numerical cartoon, $T_{\mathrm{BG}} \sim 300 $ K is room temperature.

The final dramatic change arises at temperature $T_6$ (about 800 K in our cartoon), where the phonons dominate the sound mode, which now propagates at its high temperature velocity of \begin{equation}
    v_{\mathrm{s}} = \frac{v_{\mathrm{ph}}}{\sqrt{d}}.
\end{equation}  
At temperature $T_7$ (about 1800 K in our cartoon), the heat current becomes approximately coherent; at temperature $T_8$ (about 4000 K in our cartoon) the energy current becomes coherent.  Once the energy current becomes coherent, the hydrodynamics of the coupled electron-phonon fluid becomes essentially a nearly charge neutral relativistic fluid \cite{hartnoll2018holographic}: the energy current and momentum density are approximately equivalent, and the charge current is incoherent while the energy/heat currents are nearly coherent.  Interestingly enough, the speed of sound of the fluid is also compatible with a conformal fluid, since at high temperatures $v_{\mathrm{ph}}$ plays the role of an approximate speed of light for the dominant species.  However, the fluid has the largest bulk viscosity at these high temperatures, so it is not ultimately a conformal fluid, even approximately \cite{Baier:2007ix}.  

In Section \ref{sec:kinetic}, we summarize the kinetic theory approach that we use to derive electron-phonon hydrodynamics.  While our kinetic approach is the standard one, we have nevertheless uncovered numerous phenomna (described above) that have previously been overlooked.  We further give brief arguments that the presence of rapid phonon-phonon scattering at high temperatures does not change many qualitative features of the hybrid electron-phonon fluid.  Section \ref{sec:hydro} then derives the main results about hydrodynamics that we summarized above.

In Section \ref{sec:WF}, we review the thermoelectric transport properties of an electron-phonon fluid in the presence of momentum relaxation \cite{PhysRevB.99.085104}, in light of the 8 different temperature scales described in (\ref{eq:defT}).  Although the precise temperature scales at which the WF law fails are sensitive to the precise rates of momentum relaxation, we will see that the qualitative features of thermoelectric transport in an electron-phonon \emph{hydrodynamic} metal can look quite similar to a non-hydrodynamic metal.   We will explain why, and give a qualitative overview of how the different temperature scales in (\ref{eq:defT}) relate to features in the Lorenz ratio. Since typical electron-phonon fluids exhibit violations of the Wiedemann-Franz Law just below $T_{\mathrm{BG}}$ and many of the signatures of electron-phonon hydroynamics in materials such as $\mathrm{WP}_2$ and $\mathrm{PtSn}_4$ arises at temperatures just as $L$ dips below $L_0$, it may be reasonable to estimate many electron-phonon fluids as being in the temperature regime $T_4<T<T_{\mathrm{BG}}$.  A more careful consideration of basic experimental data in Section \ref{sec:WF} confirms this expectation.  We hope, therefore,  that many of the exotic hydrodynamic phenomena described above could (in principle) be observable.  For example, in thin films, the exotic sound speed might be observable by careful studies of the temperature dependence of the real part of plasmon dispersion relations, as we will describe in Section \ref{sec:plasmons}. The imaginary part of the plasmon dispersion relation also will obtain (in two dimensions) a highly unusual imaginary part \cite{PhysRevB.93.245153} arising from the large incoherent conductivity.

In Section \ref{sec:mag}, we briefly describe hydrodynamics and transport of the electron-phonon fluid in a background magnetic field.  For simplicity we focus on the case of two dimensional systems. Depending on microscopic details of phonon-phonon scattering, the relationship between shear viscosity and Hall viscosity can either mirror a conventional Fermi liquid with negligible electron-phonon scattering \cite{scaffidi} or appear rather unconventional.

We also note there is an extensive and decades-old literature on the hydrodynamics of \emph{phonons alone}, including the propagation of hydrodynamic ``second sound" and other hydrodynamic phenomena of interacting phonons \cite{Jackson1970,Narayanamurti1972,Koreeda2007,Huberman2019,phononnat2015,lee2019,Machida309}.  This phonon-only hydrodynamics typically arises in electrical insulators.  We do not expect that the model of this paper is directly applicable to such systems, if there is a small number of thermally excited electrons or holes, because our model assumes that the electrons are non-degenerate, and that $T\ll T_{\mathrm{F}}$.  It could be interesting to generalize our results to study this alternative regime.

\section{Kinetic theory}\label{sec:kinetic}
In this section, we review a general formalism to solve the transport problem.  Our notation follows \cite{PhysRevB.97.245128,PhysRevLett.120.116603,PhysRevB.97.045105}.

\subsection{Formalism}
Consider a Fermi liquid with weak interactions.  For the moment, let us imagine a single species of particle -- we will relax this shortly.  We can describe transport by determining the response of the distribution function $f(\b{x},\b{p})$, which can roughly be interpreted as the number of quasiparticles of momentum $\b{p}$ near the spatial point $\b{x}$.  Since these quasiparticles are assumed to be long-lived, this interpretation is sensible.  The evolution of $f$ is governed by the Boltzmann equation \cite{kamenev_2011}:
\be
\pa_t f+\b{v_p}\cdot \pa_{\b{x}}f +\b{F}_{\text{ext}}\cdot \pa_{\b{p}}f=\mC[f],
\label{eq:boltzeqn}
\ee
where \begin{equation}
    \b{v_p}=\pa_{\b{p}}\ep_{\b{p}}
\end{equation} is the quasiparticle velocity, $\b{F}_{\text{ext}}$ is the external force, and $\mC[f]$ accounts for the multi-particle scattering events arising due to interactions between quasiparticles. We assume the electron band $\ep_{\b{p}}$ to be inversion symmetric and time reversal symmetric and neglect the spin degree of freedom for simplicity. We write 
\be
f(\b{x},\b{p})=f_0(\b{x},\b{p})+\d f(\b{x},\b{p}) \equiv f_0(\b{x},\b{p})+\left(-\frac{\pa f}{\pa\ep}\right)\Phi(\b{x},\b{p}),
\label{eq:linearize}
\ee
where $f_0$ is the equilibrium distribution function, and $\d f$ is the infinitesimal correction due to the deviation from equilibrium. Within linear response,  \eqnref{eq:boltzeqn} reduces to 
\be
\pa_t \d f+\b{v_p}\cdot \pa_{\b{x}}\d f +\d \b{F}_{\text{ext}}\cdot \pa_{\b{p}}f_0=\d \mC[\d f]
\label{eq:linboltzeqn}
\ee
where the right hand side is, in general, a non-local function in $\b{p}$. The external force is given by 
\be
\d \b{F}_{\text{ext}}=-e\d \b{E}+(\ep-\mu)\frac{\nabla \d T}{T}
\ee
and arises due to external electric fields and temperature gradients. 

We interpret $\Phi(\b{x},\b{p})$ as a vector, with the $\b{p}$ dependence abstracted into Dirac bra-ket notation.  Hence, we write \begin{equation}
    |\Phi\rangle = \int \mathrm{d}^dp ~\Phi(\b{x},\b{p})|\b{p}\rangle,
\end{equation} 
and define the inner product
\be
\la \b{p}|\b{p}'\ra = \left.\left(-\frac{\pa f}{\pa \ep}\right)\right|_{\b{p}} \frac{\d(\b{p}-\b{p}')}{(2\pi\hbar)^d}.
\ee
The charge and thermal current, when evaluated on a given distribution function $f$, can be written as inner products
\bes
\be
J_i(\b{x})=\la\Phi|J_i\ra=-e \int \frac{\ud^d p}{(2\pi\hbar)^d} v_i \left(-\frac{\pa f}{\pa \ep}\right)\Phi (\b{x},\b{p}),\quad |J_i\ra\equiv -e \int \ud^d p~ v_i(\b{p})|\b{p}\ra,
\ee
\be
Q_i(\b{x})=\la\Phi|Q_i\ra= \int \frac{\ud^d p}{(2\pi\hbar)^d} (\ep-\mu)v_i \left(-\frac{\pa f}{\pa \ep}\right)\Phi (\b{x},\b{p}),\quad |Q_i\ra\equiv \int \ud^d p~ (\ep-\mu)v_i(\b{p})|\b{p}\ra,
\ee
\ees
since in equilibrium there is no charge or heat current.  
We define the linearized collision integral $W$ to be the map in the vector space $W:|\b{p}\ra\to |\b{p}'\ra$ giving the linearized collision integral \begin{equation}
    \d\mC_{\b{p}}=\la \b{p}|W|\Phi\ra. 
\end{equation}The linearized Boltzmann equation becomes
\be
\pa_t|\Phi\ra + \b{v_p}\cdot \partial_{\b{x}} |\Phi\rangle -E_i|J_i\ra+\frac{\nabla_i T}{T}|Q_i\ra=-W|\Phi\ra.
\ee
We remind readers that the equilibrium distribution for the fermionic electronic quasiparticles is
\be
f_{\mathrm{F}}^0(\b{p})=\frac{1}{1+\mathrm{e}^{\beta(\ep(\b{p})-\mu)}},
\ee
while for the bosonic acoustic phonons it is
\be
f_{\mathrm{B}}^0(\b{p})=\frac{1}{\mathrm{e}^{\beta\omega(\b{p})}-1}.
\ee
The velocity of the electrons is \begin{equation}
    \b{v}_{\b{p}}=\pa_{\b{p}}\ep_{\b{p}}=v_{\mathrm{F}} \frac{\b{p}}{|\b{p}|}
\end{equation} and the velocity of the phonons is \begin{equation}
    \b{v}_{\b{p}}=\pa_{\b{p}}\omega(\b{p})=v_{\ph}\frac{\b{p}}{|\b{p}|}. \label{eq:phonondispersion}
\end{equation}


\subsection{Thermodynamics}
We assume that the system is rotationally invariant, i.e. both the dispersion relation and collision integral are rotationally invariant. Then, a convenient basis could be applied
\be
|\tilde{n},\b{m}\ra_{\e}=\int \ud^d p ~(p-p_{\mathrm{F}})^n \mathrm{Y}_{\b{m}}(\th_1,...,\th_{d-1})|\b{p}\ra,\quad |\tilde{n},\b{m}\ra_{\ph}=\int \ud^d q ~ q^n\mathrm{Y}_{\b{m}}(\th_1,...,\th_{d-1})|\b{q}\ra,
\label{eqn:tildebasis}
\ee
where $\mathrm{Y}_{\b{m}}$ is the spherical harmonics with ``angle" $\b{m}=(m_1,...,m_{d-1})$.
In the above equation, and henceforth, $p=|\b{p}|$, $\th_1,...,\th_{d-1}$ indicate the angular coordinates of $\b{p}$, and we will use two-dimensional spherical harmonic $\mathrm{Y}_m=e^{\i m \th}$  for illustration in the rest of this subsection.  The generalization to higher dimension is straightforward. The ``angle'' indices are omitted when indicating typical relaxation rate from then on. However, these basis vectors are not normalized in the radial direction, so we use the standard Gram-Schmidt method to obtain an orthonormal basis $|n,m\ra_{\e,\ph}$, denoted with no tildes. The most important normalization factors are listed below (see \appref{app:norm}):
\begin{subequations}\label{eqn:norm}
\begin{align}
\la\tilde{0},m'|\tilde{0},m\ra_{\e}&=\nu\d_{m,m'}, \\ \la\tilde{1},m'|\tilde{1},m\ra_{\e}&=\nu a^2\d_{m,m'},\\ \la\tilde{1},m'|\tilde{1},m\ra_{\ph}&=2w^2\d_{m,m'}.
\end{align}
\end{subequations}
In our example of $d=2$, we can write the total momentum, the charge current and the thermal current as 
\bes
\be
|p_x\ra\pm\i|p_y\ra=p_{\mathrm{F}}|\tilde{0},\pm1\ra_{\e}+|\tilde{1},\pm1\ra_{\e}+|\tilde{1},\pm1\ra_{\ph},
\ee
\be
|J_x\ra\pm\i|J_y\ra=-e(v_{\mathrm{F}}|\tilde{0},\pm1\ra_{\e}+\pa_pv_{\mathrm{F}}|\tilde{1},\pm1\ra_{\e}+...),
\ee
\be
|Q_x\ra\pm\i|Q_y\ra=v_{\mathrm{F}}^2|\tilde{1},\pm1\ra_{\e}+...+v_{\ph}^2|\tilde{1},\pm1\ra_{\ph}.
\ee
\ees
The momentum operator is exact while the current operators are only written to  leading order in $T/T_{\mathrm{F}}$. Besides, the energy density vector $|\ep\ra$ and charge density vector $|\rho\ra$ are given by
\begin{equation}
    |\ep\ra=\mu|\rho\ra+|\tilde{\ep}\ra=\mu|\tilde{0},0\ra_{\e}+v_{\mathrm{F}}|\tilde{1},0\ra_{\e}+v_{\ph}|\tilde{1},0\ra_{\ph}+...,
\end{equation}
where $|\tilde{\ep}\ra$ is the energy density part orthogonal to density. Based on above identities, important thermodynamic properties including momentum susceptibility $M$ (mass density in a Galilean-invariant fluid), specific heat $c$,  charge density $\rho$ and energy density $\epsilon$ could be calculated explicitly, 
\bes
\be
M\equiv\la p_x|p_x\ra=\frac{\nu}{2}p_{\mathrm{F}}^2\left(1+\ba^2+\bw^2\right),
\ee
\be
c = T\pa_T \ep=\la \tilde{\ep}|\tilde{\ep}\ra=\frac{\nu}{2}v_{\mathrm{F}}^2p_{\mathrm{F}}^2\left(\ba^2+r^2\bw^2+...\right),
\label{eq:specificheat}
\ee
\be
\rho\equiv\frac{\la J_x|p_x\ra}{-e}=\frac{\nu}{2}v_{\mathrm{F}}p_{\mathrm{F}}\left(1+(p_{\mathrm{F}}\pa_p\ln v_{\mathrm{F}}) \ba^2+...\right),
\label{eq:chden}
\ee
\be
\epsilon \equiv\frac{\la Q_x|p_x\ra}{T}=\frac{\nu}{2T}v_{\mathrm{F}}^2p_{\mathrm{F}}^2\left(\ba^2+r^2\bw^2+...\right).
\label{eq:enden}
\ee
\ees


\subsection{Electron-phonon collision integral}\label{sec:e-ph}
We now describe the collision integerals, beginning with the electron-phonon interactions.  The dominant such interaction is a single phonon emission/absorption event \cite{levchenko2020transport,RevModPhys.89.015003}:
\be
H_{\text{e-ph}}=\sum \L_{\b{q},\b{k}_1,\b{k}_2}(a_{\b{q}}+a_{-\b{q}}^\dagger)c^\dagger_{\b{k}_1}c_{\b{k}_2}, \label{eq:displacement}
\ee
The Boltzmann equations for electrons and phonons are
\begin{subequations}
\begin{align}
\pa_t f_{\mathrm{F}} +\b{v}_{\mathrm{F}}\cdot \pa_{\b{x}}f_{\mathrm{F}} +\b{F}\cdot \nabla_{\b{p}} f_{\mathrm{F}}&=\mC_{\text{e-ph}},\\
\pa_t f_{\mathrm{B}} +\b{v}_{\ph}\cdot \pa_{\b{x}}f_{\mathrm{B}} +\b{F}\cdot \nabla_{\b{p}} f_{\mathrm{B}}&=\mC_{\text{ph-e}},
\end{align}
\label{eq:ephboltz}
\end{subequations}
where 
\be
\mC_{\text{e-ph}}=\int \mathrm{d}^d\b{q} \mathrm{d}^d\b{k}_1 |\L|^2\d(\b{k}_2-\b{k}_1-\b{q})\d(\ep_{k_2}-\ep_{k_1}-\omega_q)\left\{f_{\mathrm{F}k_1}(1-f_{\mathrm{F}k_2})f_{\mathrm{B}q}-f_{\mathrm{F}k_2}(1-f_{\mathrm{F}k_1})(1+f_{\mathrm{B}q})\right\}.
\label{e-ph-coll}
\ee
$\mC_{\text{ph-e}}$ is given by the same equation but with an integral over $\b{k}_1,\b{k}_2$. In the long-wavelength limit, the transition probability is approximated by \cite{ziman2001electrons} 
\begin{equation}
    |\Lambda^2|\sim D_{\text{e-ph}}|\b{q}|.
\end{equation}
Following \eqnref{eq:linearize}, we are able to linearize the Boltzmann equation (see explicit derivation in \appref{app:e-ph})
\be
\la\Phi| W_{\text{e-ph}} |\Phi\ra=\beta\int \mathrm{d}^d\b{q} \mathrm{d}^d\b{k}_1\mathrm{d}^d\b{k}_2|\L|^2\d(\b{k}_2-\b{k}_1-\b{q})\d(\ep_{k_2}-\ep_{k_1}-\omega_q)(1-f_{\mathrm{F}k_2})f_{\mathrm{F}k_1}f_{\mathrm{B}q}|\Phi_{k_1}-\Phi_{k_2}+\Phi_{q}|^2.
\label{S}
\ee
However, after linearization, \eqnref{eq:ephboltz} are still coupled integrodifferential equations which are difficult to solve analytically. Nevertheless, a typical relaxation rate is good enough to estimate the scalings of overall prefactors of hydrodynamics and thermodynamic properties of thermoelectric transport. To obtain the typical scattering rate for electron (phonon) modes, we set the Ansatz $\Phi_{\ph}=0$ ($\Phi_{\e}=0$). We summarize the results here, and refer readers to the explicit calculations in \appref{app:e-ph}:
\bes
\begin{equation}
    \ga_{\e}=\la 0|W_{\text{e-ph}}|0\ra_\e=
    \left\{ \begin{array}{ll}
     \ga & T<T_{\text{BG}} \\
     \bar{\gamma}  & T_{\text{BG}}<T\ll T_{\mathrm{F}}
\end{array} \right. ,
\end{equation}
\begin{equation}
    \ga_{\e}^{\prime}=\la 1|W_{\text{e-ph}}|1\ra_\e=\left\{ \begin{array}{ll}
     \ga r^2/\ba^2 & T<T_{\text{BG}} \\
     \bar{\gamma}  & T_{\text{BG}}<T\ll T_{\mathrm{F}}
\end{array} \right. ,
\end{equation}
\begin{equation}
    \ga_{\ph}=\la 1|W_{\text{e-ph}}|1\ra_\ph=\left\{ \begin{array}{ll}
     \ga /\bw^2 & T<T_{\text{BG}} \\
     \bar{\gamma}/\bw^2  & T_{\text{BG}}<T\ll T_{\mathrm{F}}
\end{array} \right. ,
\end{equation}
\ees
where $\ga$ and $\bar{\gamma}$ are celebrated electron-phonon scattering rates seperated by Bloch-Gr$\ddot{\text{u}}$neisen temperature \cite{PhysRevLett.105.256805}:
\bes\label{eqn:gamma}
\begin{equation}
    \gamma=\a^2(2) \frac{2}{\nu p_{\mathrm{F}}^2}T^{d+2}\sim \gamma_0\left(\frac{T}{T_{\mathrm{BG}}}\right)^{d+1}\left(\frac{T}{T_{\mathrm{F}}}\right)\propto T^{d+2},
\end{equation}
\begin{equation}
    \bar{\gamma}=\a'^2(2) \frac{2}{\nu p_{\mathrm{F}}^2}T\sim \gamma_0 \left(\frac{T}{T_{\mathrm{F}}}\right) \propto T,
\end{equation}
\ees
where
\begin{equation}\label{eqn:gamma0}
    \gamma_0=D_{\text{e-ph}}\frac{p_{\mathrm{F}}^d}{v_{\ph}}.
\end{equation}
So far, the results are exact, and they consist of diagonal terms of the collision integral. To account for the momentum conservation, we multiply the collision integral with projectors:
\be
W_{\text{e-ph}}'=(\mathbbm{1}-\bbP)W_{\text{e-ph}}(\mathbbm{1}-\bbP),\quad \bbP=\frac{|p_x\ra\la p_x|}{\la p_x|p_x\ra}.
\label{eqn:Weph}
\ee
The explicit expression can be found in \appref{app:wf} by ignoring impurity scattering there. The projector above has two physical meanings: first, it makes the total momentum a null vector of the collision integral, i.e. the total momentum has no relaxation; second, the projector gives (approximately) off-diagonal terms of the collision integral by mixing the diagonal terms, taking into account the fact that either electron or phonon momentum could be transferred to each other through electron-phonon scattering even if the total momentum is conserved. Nevertheless, thanks to the semi-positivity of the collision integral (given directly by the second-law of thermodynamics), the off-diagonal terms cannot be qualitatively important, that is, using diagonal terms is good enough to determine the scaling in hydrodynamics.

\subsection{Phonon-phonon collision integral}\label{sec:ph-ph}
When the temperature is greater than the Bloch-Gr$\ddot{\text{u}}$neisen temperature $T_{\text{BG}}$, high momentum phonon modes $q>p_{\mathrm{F}}$ are allowed to appear in electron phonon fluid. However, if there is only electron-phonon interaction, they effectively cannot be scattered via the coupling in \eqnref{eq:displacement} due to the simultaneous momentum conservation and energy conservation (no two electrons can be connected by such a large momentum transfer). A priori, this might seem to lead to the conclusion that there is some non-interacting part of phonon modes that does not participate in hydrodynamics, at least until higher order scattering processes are considered.  Another practical solution is simply to account for phonon-phonon scattering processes.  For simplicity, we also ignore umklapp phonon-phonon processes, so that momentum remains exactly conserved in the absence of impurities.   
In what follows, we analyze the phonon-phonon interaction schematically.  Our primary goal is to show that the phonon-phonon interaction will couple high energy phonons to low energy phonons, and hence not lead to any additional long-lived degrees of freedom in the hydrodynamic description. 

Due to the relativistic dispersion relation of acoustic phonon modes, momentum and energy conservation together imply fast collinear scattering \cite{PhysRevB.94.205306,Lucas_2018}.  To see this explicitly, we take a ``random" ansatz in phonon subspace
\be
|\Phi\ra=\int \ud^d q|\b{q}\ra_{\ph}, \quad \la \Phi|\Phi\ra\sim r^d\left(\frac{T}{T_{\mathrm{F}}}\right)^{d-1}.
\ee
The first order phonon-phonon interaction 
\be
H_{\text{ph-ph}}=\sum_{\b{q},\b{k}}U(\b{q},\b{k})a^\dagger_{\b{q}+\b{k}}a_{\b{q}}a_{\b{k}}+\text{h.c.}
\ee
gives rise to the linearized collision integral under the ``random" Ansatz 
\be
\beal
\la \Phi|W_{\text{ph-ph}}|\Phi\ra&=\int \ud^dq_1\ud^dq_2\ud^dq_3 |U|^2\d(\b{q}_1-\b{q}_2-\b{q}_3)\d(|\b{q}_1|-|\b{q}_2|-|\b{q}_3|)\frac{1}{v_{\ph}}\beta f_{\mathrm{B}q_1}(1+f_{\mathrm{B}q_2})(1+f_{\mathrm{B}q_3})\\
&=\int^{T/v_{\ph}} \ud^dq_2\ud^dq_3\d(|\b{q}_2+\b{q}_3|-|\b{q}_2|-|\b{q}_3|)\frac{T^{2}D_{\ph}}{v_{\ph}^4},
\eeal
\label{eqn:phph}
\ee
where we have noted that the phonon-phonon interaction typically takes the form \cite{ziman2001electrons} 
\begin{equation}
    |U|^2\sim D_{\ph}|\b{q}_1||\b{q}_2||\b{q}_3|.
\end{equation}
Obviously, the $\d$-function is only satisfied when the momentums stay parallel, i.e. realizing collinear scattering. Schematically, we denote the high momentum phonon modes ($q>p_{\mathrm{F}}$) and low momentum phonon modes ($q<p_{\mathrm{F}}$) to be $|\text{H}\ra_{\ph}$ and $|\text{L}\ra_{\ph}$ (normalized). Since collinear scattering preserves momentum, it cannot relax the phonon momentum,  
but can mix $|\text{L}\ra_{\ph}$ and $|\text{H}\ra_{\ph}$ effectively. We identify the typical relaxation rate for $|\text{H}\ra_{\ph}$ to relax to $|\text{L}\ra_{\ph}$ as the collinear scattering rate $\ga_{\text{ph-ph,coll}}=\la \Phi|W_{\text{ph-ph,coll}}|\Phi\ra/\la\Phi|\Phi\ra$. 
Let 
\be
\b{q}_2=(q_2,0),\quad \b{q}_3=(q_3,\b{q}_{\perp})
\ee
where we assume $\b{q}_{2\perp}=0$ and $\b{q}_{3\perp}\equiv \b{q}_{\perp}\ll q_{2,3}$ for collinear scattering. 
Then the delta function in \eqnref{eqn:phph} becomes
\be
\d(|\b{q}_2+\b{q}_3|-|\b{q}_2|-|\b{q}_3|)=2\d\left(\frac{q_{\perp}^2}{q_2+q_3}-\frac{q_{\perp}^2}{q_3}\right)=\frac{2(q_2+q_3)q_3}{q_2}\d(q_{\perp}^2).
\ee
The collision integral is estimated through
\be
\la \Phi|W_{\text{ph-ph,coll}}|\Phi\ra=\int^{T/v_{\ph}}  \ud q_2\ud q_3 \frac{2(q_2+q_3)q_3}{q_2}\frac{T^2D_{\ph}}{v^4_{\ph}}\int \ud^{d-1}q_{\perp}q^{-1}_{\perp}\d(q_{\perp})\sim r^{-7}\left(\frac{T}{T_{\mathrm{F}}}\right)^5\times \left\{ \begin{array}{ll}
  \log(1/\a)& d=2 \\
 \text{const.}& d>2
\end{array} \right. ,
\ee
where the integral has a logarithm divergence in $d=2$ \cite{PhysRevB.94.205306} with $\a \to 0$ the cutoff from self-energy correction \cite{PhysRevB.78.085416}. It is known as collinear scattering singularity \cite{Lucas_2018}, which is not truly divergent as long as the scattering amplitude $|U|^2$ vanishes for collinear scattering. Then the collinear scattering rate is given by $\ga_{\text{ph-ph,coll}}\approx T\times A_{\text{ph-ph,coll}}$ with 
\be
A_{\text{ph-ph,coll}}\approx D_{\ph} \frac{v_\ph^{d-7}}{T^{d-5}}
\ee
being the dimensionless parameter characterizing the typical amplitude of the collinear phonon-phonon scattering. For $T>T_{\text{BG}}$, we can also write the electron-phonon scattering rate for phonon modes as $\ga_{\text{e-ph}}\approx T\times A_{\text{e-ph}}$ with
\be
A_{\text{e}}\approx D_{\text{e-ph}}\frac{2p_{\mathrm{F}}^{2(d-1)}}{\nu v_{\mathrm{F}}^2 }\frac{1}{v_{\ph}},\quad A_{\text{ph}}\approx D_{\text{e-ph}} \frac{p_{\mathrm{F}}^{2d}}{v_{\mathrm{F}}^2} \frac{v_\ph^{d+1}}{T^{d+1}},
\ee
and we find $A_{\text{ph-ph,coll}}\gg A_{\text{e-ph}}$ at $T>T_{\text{BG}}$, assuming that $D_{\text{e-ph}}$ and $D_{\ph}$ are similar magnitude and approximately momentum-independent. It suggests that due to the phonon-phonon interaction, the rate for $|\text{H}\ra_{\ph}$ to relax to $|\text{L}\ra_{\ph}$ is much faster than that for the $|1\ra_{\ph}$ (effectively $|\text{L}\ra_{\ph}$) to relax out due to electron-phonon interaction. In another word, the total electron and phonon momentum remains a good conserved quantity at the time scale of $\ga_{\text{e-ph}}^{-1}$, thus the electron-phonon fluid is well-defined at $T>T_{\text{BG}}$.
For concreteness, let's write the momentum as
\begin{equation} 
     |P\rangle \propto  c_2 |0\rangle_{\e} + c_1c_2 |\text{L}\rangle_{\ph} + c_1 |\text{H}\rangle_{\ph},
     \label{eqn:p-coll}
\end{equation}
where $c_{1,2}$ are dimensionless constants. The collision integral with momentum conservation is of the form
\begin{equation}
        \langle \Phi|W|\Phi\rangle = \gamma_{\text{e-ph}}\frac{(\Phi_{\text{L}}-c_1\Phi_{\e})^2}{1+c_1^2} + \gamma_{\text{ph-ph,coll}}\frac{(\Phi_{\text{L}}-c_2\Phi_{\mathrm{H}})^2}{1+c_2^2}.
\end{equation}
In the limit of $\gamma_{\text{e-ph}}/\gamma_{\text{ph-ph,coll}}\ll (1+c_1^2)(1+c_2^2)$, the eigenvalues of $W$ are given by $\{0,\gamma_{\text{e-ph}},\gamma_{\text{ph-ph,coll}}\}$ approximately. The null eigenvalue comes from momentum conservation. Notice that we obtain a looser constraint on $\gamma_{\text{e-ph}}/\gamma_{\text{ph-ph,coll}}$ since $c_1>1$ at high temperature. The eigenstate, corresponding to the eigenvalue $\gamma_{\text{ph-ph}}$, is $c_1(1+c_1^2)^{-1}(\gamma_{\text{e-ph}}/\gamma_{\text{ph-ph,coll}})|0\ra_{\e}-|\text{L}\ra_{\ph}+c_2|\text{H}\ra_{\ph}$, so that, in accordance with \eqnref{eqn:p-coll}, electron modes are strongly suppressed in phonon-phonon collinear scattering, coinciding with the previous argument.

Observe that the modes $|\tilde{1},m\rangle_{\ph}$ (which represent the phonon momentum) are exact null vectors of the linearized phonon-phonon collision integral for $m=0$ due to conservation of phonon energy in phonon-phonon collisions, and when $m=\pm 1$ due to conservatioon of momentum.  Once $|m|>1$, $|\tilde{1},m>1\ra_{\ph}$, will scatter non-collinearly and experience a non-collinear phonon-phonon scattering rate
\be
\ga_{\text{ph-ph,non-coll}}\sim \frac{D_{\ph}}{\la \Phi|\Phi\ra} \left(\frac{T}{v_{\ph}}\right)^{2d-1} \frac{T^2}{v_{\ph}^4} \sim D_{\ph} \frac{T^{d+2}}{v_{\ph}^{d+3}} .
\ee
Formally, in the hydrodynamic calculations, the collision integral becomes
\be
W^{\prime}=W_{\text{e-ph}}^{\prime}+(\ga_{\text{ph-ph,non-coll}}+\ga_{\text{ph-ph,coll}}\left(\mathbbm{1}-\bbP_{\text{L,H}}\right))\sum_{|m|>1} \bbP_{m}
\label{eq:noncoll}
\ee
where $\bbP_{m}=|1,m\ra_{\ph}\la1,m|_{\ph}$ is the normalized projector onto phonon modes with ``angle" $m$, and $\bbP_{\text{L,H}}$ is the projection within phonon sector making phonon momentum a null vector to the collision integral. 

Results of scattering rates computed in \secref{sec:e-ph} and \secref{sec:ph-ph} are summarized in \tabref{tab:rate}.
\begin{table}[t]
\begin{tabular}{|c|c|c|c|c|}
\hline
$T$ & $<T_3$ & $T_3$ to $T_4$ & $T_4$ to $T_{\text{BG}}$ & $T_{\text{BG}}$ to $T_{\mathrm{F}}$\\
\hline
scalings  &   \multicolumn{3}{|c|}{$\ga_{\e}=\ga\sim T^{d+2}, ~\gamma^{\prime}_{\e}=\ga r^2/\ba^2\sim v_{\ph}^2T^{d}, ~\ga_{\ph}=\ga/\bw^2\sim v_{\ph}^{d+2}T$} & $\ga_{\e}=\gamma^{\prime}_{\e}=\bar{\gamma}\sim T,~ \ga_{\ph}=\bar{\ga}/\bw^2\sim v_{\ph}^{d+2}T^{-d}$ \\
\hline
relations & $ \ga_{\e}<\gamma^{\prime}_{\e}<\ga_{\ph}$ & $ \ga_{\e}<\ga_{\ph}<\gamma^{\prime}_{\e}$ & $ \ga_{\ph}<\ga_{\e}<\gamma^{\prime}_{\e}$  & $ \ga_{\ph}<\ga_{\e}=\gamma^{\prime}_{\e}\ll \ga_{\text{ph-ph}}$ \\
\hline
\end{tabular}
\caption{Summary of scattering rate in different temperature regimes. $\ga_{\text{ph-ph}}$ is a short hand for phonon-phonon scattering rate.}
\label{tab:rate}
\end{table}




\section{Hydrodynamics}\label{sec:hydro}


Having developed the kinetic theory of electron phonon fluid, we can study its universal properties in the hydrodynamic limit of $\pa_t \ll \min W_{\text{e-ph}}^{\prime}$. Hydrodynamics is an effective theory approach and describes the late time dynamics of the conserved quantities.  In our rotationally invariant electron-phonon fluid, there are charge, energy and momentum: \begin{subequations} \label{eq:hydromain}
\begin{align}
    \pa_t \rho +\pa_i J_i&=0, \\
    \pa_t \ep +\pa_i J_{\mathrm{E}i}&=0, \\
    \pa_t p_i +\pa_j \tau_{ij}&=0,
\end{align}
\end{subequations}
and
\be
\beal
\left(\begin{array}{c}
J_i \\
J_{\mathrm{E}i} \end{array}\right)&=\left(\begin{array}{c}
\rho \\
\ep+P \end{array}\right)u_i-\Sigma_0\pa_i
\left(\begin{array}{c}
\mu \\
T \end{array}\right)+
\cdots\\
\tau_{ij}&=-\eta(\pa_iu_j+\pa_ju_i-\frac{2}{d}\d_{ij}\pa_ku_k)-\zeta\d_{ij}\pa_ku_k+\d_{ij}P+...
\eeal
\label{visc}
\ee
where $\Sigma_0$ is the incoherent conductivity matrix \cite{Hartnoll2015,hartnoll2018holographic,Davison:2015taa}, $\eta$ is the shear viscosity and $\zeta$ is the bulk viscosity. 
The hydrodynamic quasinormal modes, which are the degrees of freedom of the effective theory, are found by substituting a plane wave ansatz into (\ref{eq:hydromain}). We obtain a sound mode
\be
\omega=\pm v_{\mathrm{s}} k-\i \Gamma k^2+\mO(k^3)
\ee
where $v_{\mathrm{s}}$ is the sound velocity, and $\Gamma$ is the sound wave's ``diffusion constant". Apart from the sound mode, the perpendicular components of the velocity field (the transverse momentum) obey a diffusion equation controlled by the shear viscosity:
\begin{equation}
    \omega=-\i \frac{\eta}{M}k^2.
\end{equation}
For convenience, we introduce operators: $\rho_A=(\rho,\ep), \mu_A=(\mu,T), A=1,2$. Assisted with kinetic formalism, we could calculate explicitly the transport coefficients in hydrodynamic equations in following subsections.

\begin{figure}[t]
 \includegraphics[width=1.\linewidth]{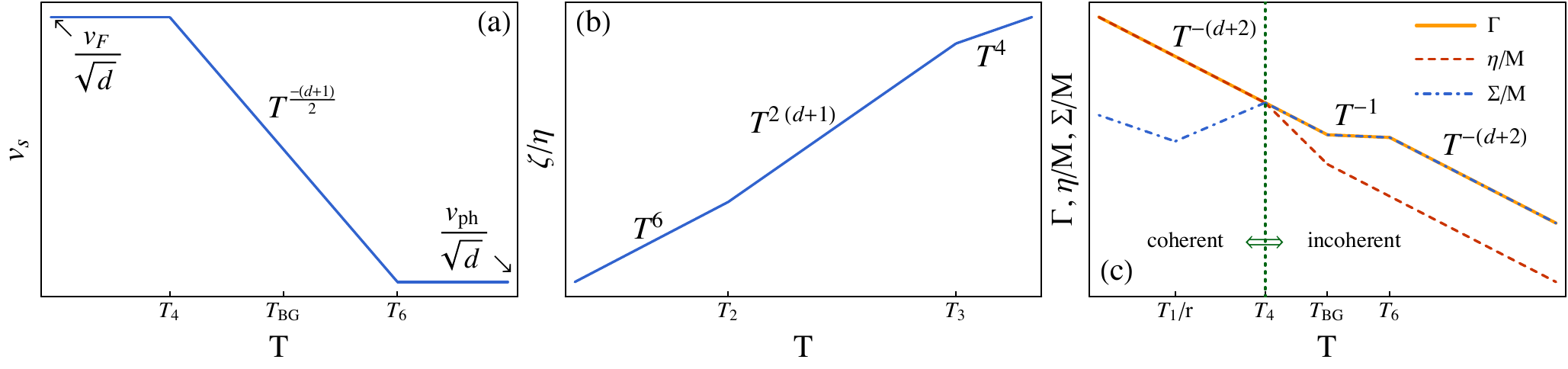}
\caption{Hydrodynamics of electron-phonon fluid. (a): Log-log plot of sound velocity against temperature. (b): Log-log plot of bulk viscosity over shear viscosity against temperature. (c): Log-log plot of diffusion constant (orange thick line), shear viscosity (red dashed line) and incoherent conductivity (blue dotdashed line) against temperature. In (c), the vertical green line separates the fluid into coherent and incoherent. }
\label{fig:hydro}
\end{figure}

\subsection{Speed of sound}

Based on \eqnref{eq:chden} and \eqnref{eq:enden}, we can derive the susceptibility matrix
\begin{align}
\chi\equiv \frac{\pa \rho_A}{\pa \mu_B}=
\left(\begin{array}{cc}
\frac{\pa \rho}{\pa \mu} &\frac{\pa \rho}{\pa T}  \\
\frac{\pa \ep}{\pa \mu}& \frac{\pa \ep}{\pa T}
\end{array}\right)\approx 
\left(\begin{array}{cc}
\frac{d \nu}{2} & \frac{\pa}{\pa T}(\frac{\nu}{2}\pa_p v_{\mathrm{F}}a^2) \\
\frac{\pa}{\pa \mu}(\frac{\nu v_{\mathrm{F}}^2a^2}{2T})& \nu(\frac{\pi^2}{6}+I^2\frac{dT^{d-1}}{v_{\mathrm{ph}}^d\nu})
\end{array}\right)\Rightarrow
\chi^{-1}\approx 
\left(\begin{array}{cc}
 \frac{2}{d\nu } &\mO(\frac{T}{T_{\mathrm{F}}}) \\
\mO(\frac{T}{T_{\mathrm{F}}})& \frac{d\nu}{2\det \chi}
\end{array}\right),
\end{align}
where \begin{equation}
    \det \chi\approx \frac{d}{2} \left(\frac{\pi^2}{6}\nu^2+I^2\frac{dT^{d-1}}{v_{\mathrm{ph}}^d}\nu\right).
\end{equation} 
Note that $\det\chi$ switches from being dominated by electrons to dominated by phonons at $T\sim T_3$.
Keeping the leading order in $T/T_{\mathrm{F}}$, the sound wave velocity of the quasinormal mode is given by
\be
\beal
v_{\mathrm{s}}=&\sqrt{\frac{\rho_A\chi^{-1}_{AB}\rho_B}{M}}\approx\sqrt{\left(\frac{1}{d}\nu^2\mu^2+2d\left(\frac{\nu^2}{2}v_{\mathrm{F}}^2p_{\mathrm{F}}^2\right)^2\frac{(\ba^2+r^2\bw^2)^2}{T^2\det \chi}\right)\Big/\nu^2p_{\mathrm{F}}^2\left(1+\ba^2+\bw^2\right)}\\
\approx&\left\{ \begin{array}{ll}
\displaystyle \sqrt{\frac{\mu^2}{d p_{\mathrm{F}}^2}}= \frac{v_{\mathrm{F}}}{\sqrt{d}} & T<  T_4 \\
\displaystyle \sqrt{\frac{\mu^2}{dp_{\mathrm{F}}^2\bw^2}}= \frac{v_{\mathrm{F}}}{\bw\sqrt{d}} & T_4<T<  T_6 \\
\displaystyle \sqrt{\frac{v_{\mathrm{F}}^2r^2}{d}}= \frac{v_{\mathrm{ph}}}{\sqrt{d}}& T_6 <T\ll T_{\mathrm{F}}
\end{array} \right. ,
\eeal
\label{eqn:soundv}
\ee
where repeated indices are summed. To obtain the last phonon-limited sound velocity, we have used the identity \eqnref{eqn:bw} to replace constant $I$ with dimensionless $\bw$. The scaling is shown in \figref{fig:hydro}(a). We observe that when temperature is low enough, electron phonon fluid is more like a Fermi liquid with sound velocity given by the Fermi velocity. When the temperature rises over $T_{4}$, the phonon momentum starts to dominate over the total momentum such that the sound velocity is corrected by a factor of $1/\bw<1$. 
With Fermi velocity and phonon momentum weight appearing in the same expression, this regime cannot be thought of as phonon-dominated or electron-dominated, but electrons and phonons together flow as a single unified fluid.
When the temperature is higher than $T_{6}$, the sound velocity is controlled by the phonon velocity $v_\mathrm{ph}/\sqrt{d}$.



\subsection{Shear viscosity}
To calculate the shear viscosity, we first write down the momentum current (stress tensor) operator.  Focusing on $d=2$ for ease of presentation:
\be
\beal
|\tau_{xx}\ra-|\tau_{yy}\ra&=\int \ud^d p pv_p\cos 2\th |\b{p}\ra+v_{\ph}\int \ud^d q q\cos 2\th |\b{q}\ra\\
&=\frac{1}{2}\left(p_{\mathrm{F}}v_{\mathrm{F}}|\tilde{0},2\ra_{\e}+(v_{\mathrm{F}}+p_{\mathrm{F}}\pa_pv_{\mathrm{F}})|\tilde{1},2\ra_{\e}+\pa_pv_{\mathrm{F}}|\tilde{2},2\ra_{\e}+...+v_{\ph} |\tilde{1},2\ra_{\ph}+(m=-2)\right).
\eeal
\ee
In above equation, all the modes are in $m=\pm2$ sector, thus there is no good conservation law.  We then approximate that the collision integral in this sector is diagonal, and obtain
\be
\beal
\eta=&\left(\la\tau_{xx}|-\la\tau_{yy}|\right)W^{\prime -1}\left(|\tau_{xx}\ra-|\tau_{yy}\ra\right)\\
\approx& \nu\mu^2\ga^{-1}_{\e}+w^2v_{\ph}^2(\ga_{\ph}+\ga_{\text{ph-ph}})^{-1}\\
\approx& \left\{ \begin{array}{ll}
\nu\mu^2 \ga^{-1}  & T<T_{\text{BG}}\\
\nu\mu^2 \left(\bar{\gamma}^{-1}+r^2\bw^2\ga_{\text{ph-ph}}^{-1} \right) & T_{\text{BG}}<T\ll T_{\mathrm{F}}\\
\end{array} \right. .
\eeal
\label{eq:shear}
\ee
The change in $\eta$ manifests in Bloch-Gr$\ddot{\text{u}}$neisen effect with $\gamma\to \bar{\gamma}$. Note that when $T>T_{\text{BG}}$, the shear viscosity may still be dominated by the electron modes due to the fast phonon-phonon scattering (including both collinear and non-collinear scattering).  However, it is also plausible that, depending on microscopic details, above $T>T_{\mathrm{BG}}$ phonons can dominate the shear viscosity.

\subsection{Bulk viscosity}\label{sec:bulk}
Unlike the shear viscosity, to estimate the bulk viscosity, we need to search for the non-conserved part of the trace of the momentum current.  Again writing formulas in $d=2$ for illustrative purposes, we obtain
\be
\beal
|\tau_{xx}\ra+|\tau_{yy}\ra&=\int \ud^d p pv_p |\b{p}\ra+v_{\ph}\int \ud^d q q|\b{q}\ra\\
&=p_{\mathrm{F}}v_{\mathrm{F}}|\tilde{0},0\ra_{\e}+(v_{\mathrm{F}}+p_{\mathrm{F}}\pa_pv_{\mathrm{F}})|\tilde{1},0\ra_{\e}+\pa_pv_{\mathrm{F}}|\tilde{2},0\ra_{\e}+...+v_{\ph} |\tilde{1},0\ra_{\ph}.
\eeal
\ee
To this end, we consider the incoherent momentum current 
\begin{align} \label{eq:tauxxyyinc}
(|\tau_{xx}\ra+|\tau_{yy}\ra)_{\inc}&\equiv|\tau_{xx}\ra+|\tau_{yy}\ra-\frac{\la\tilde\ep|(|\tau_{xx}\ra+|\tau_{yy}\ra)}{\la\tilde\ep|\tilde\ep\ra}|\tilde\ep\ra -\frac{\la\rho|(|\tau_{xx}\ra+|\tau_{yy}\ra)}{\la\rho|\rho\ra}|\rho\ra \notag \\
&\approx\frac{2r^2w^2}{\la\tilde{\ep}|\tilde{\ep}\ra}v_{\mathrm{F}}|\tilde{1},0\ra_{\e}-\frac{a^2}{\la\tilde{\ep}|\tilde{\ep}\ra}v_{\ph}|\tilde{1},0\ra_{\ph}+...+|\tilde{2},0\ra_{\e}.
\end{align}
While vanishing for a quadratic electron band, the $|\tilde{2},0\ra_{\e}$ contribution above is present for a generic dispersion relation. In above, we omit the temperature-independent (also $v_{\ph}$-independent) prefactor to simplify the scaling analysis. The bulk viscosity is given by 
\be
\beal
\zeta&=(\la\tau_{xx}|+\la\tau_{yy}|)_{\text{inc}}W_{\text{e-ph}}^{\prime -1}(|\tau_{xx}\ra+|\tau_{yy}\ra)_{\inc}\\
&=\nu(v_{\mathrm{F}}p_{\mathrm{F}})^2\left(\frac{r^4 \ba^2\bw^4 \ga_{\e}'^{-1}+r^2 \ba^4 \bw^2\ga_{\ph}^{-1}-2r^3\ba^3\bw^3\ga_{\e-\ph}'^{-1}}{(\ba^2+r^2\bw^2)^2}+\left(\frac{T}{T_{\mathrm{F}}}\right)^4\ga_{\e}''^{-1}\right),
\eeal
\ee
where $\ga_{\e}''=\la 2|W_{\text{e-ph}}|2\ra_\e\approx r^4\ba^{-2}\ga$ when $T<T_{\text{BG}}$ (see \appref{app:e-ph}).
After some algebra, we find that
\be
\zeta\approx\left\{ \begin{array}{ll}
 \nu(v_{\mathrm{F}}p_{\mathrm{F}})^2 \ga^{-1}r^{-4}\ba^6\sim  \eta r^{-4}(T/T_{\mathrm{F}})^6  & T<T_2\\
\nu(v_{\mathrm{F}}p_{\mathrm{F}})^2\ga^{-1} r^2\bw^4 \sim \eta   r^{-(2d+2)}(T/T_{\mathrm{F}})^{2d+2} & T_2<T< T_3\\
\nu(v_{\mathrm{F}}p_{\mathrm{F}})^2\ga^{-1} r^{-2}\ba^4 \sim \eta r^{-2}(T/T_{\mathrm{F}})^4 &
T_3<T< T_{\text{BG}} \\
\nu(v_{\mathrm{F}}p_{\mathrm{F}})^2\bar{\gamma}^{-1} r^{-2}\ba^4 \sim  \eta r^{-2}(T/T_{\mathrm{F}})^4  &
T_{\text{BG}}<T\ll T_{\mathrm{F}}\\
\end{array} \right. .
\ee
The first two critical temperatures are unique to the bulk viscosity (see \figref{fig:hydro}(b)). $T_2$ is the critical temperature below which the incoherent momentum current, i.e. the non-conserved part of the momentum current, is predominated by $|\tilde{2},0\ra_{\e}$. This is the regime where electron-phonon fluid is completely dominated by the electron modes. However, unlike a generic Fermi liquid, the relation $\zeta/\eta\sim (T/T_{\mathrm{F}})^4$ is not satisfied in such electron-phonon fluid even $T<T_2$ because $\ga_{\e}''\neq \ga_{\e}$ for electron-phonon interaction. After all, the radial deformations of the Fermi surface decay quite differently from radially uniform deformations.   The electron-phonon fluid is intrinsically different from the Fermi liquid dominated by electron-electron interactions, even if transport is dominated by the electron modes. $T_3$ is the critical temperature above which the phonon energy starts to dominate the total energy.

Note that the bulk viscosity is much smaller than the shear viscosity at low temperatures.  Interestingly, since shear viscosity is suppressed at high temperature due to the non-collinear phonon-phonon scattering, in principle the bulk viscosity can exceed the shear viscosity at sufficiently high temperature. The non-vanishing of bulk viscosity at high temperatures also implies that the relativistic electron-phonon fluid at high temperature is strongly non-conformal, as conformal symmetry fixes $\zeta=0$ \cite{Baier:2007ix}.  Note that a phonon fluid on its own, however, would have $\zeta=0$ since $(|\tau_{xx}\rangle + |\tau_{yy}\rangle)_{\mathrm{inc}}=0$, as can readily be seen from (\ref{eq:tauxxyyinc}).\footnote{This remains true until we consider subleading $p^2$ corrections to the acoustic phonon dispersion relation (\ref{eq:phonondispersion}).}


\subsection{Incoherent conductivities}\label{sec:inc}
To calculate the incoherent conductivities, we need first to compute the incoherent currents. The incoherent charge current is given by
\be
\beal
\norm{|J_{\inc}\ra}&\equiv\norm{|J_x\ra-\frac{\la p_x |J_x\ra}{\la p_x|p_x\ra}|p_x\ra}
\approx\sqrt{\frac{e^2\nu}{2}}\sqrt{ (v_{\mathrm{F}}^2+p_{\mathrm{F}}^2\pa_pv_{\mathrm{F}}^2 \bar{a}^2)-\frac{(v_{\mathrm{F}}+p_{\mathrm{F}}\pa_pv_{\mathrm{F}}\bar{a}^2)^2}{1+\ba^2+\bar{w}^2}}\\
&\approx \left\{ \begin{array}{ll}
 \sqrt{\nu/2}e |p_{\mathrm{F}}\pa_pv_{\mathrm{F}}-v_{\mathrm{F}}|\ba \sim T/T_{\mathrm{F}} & T<  T_1 \\
 \sqrt{\nu/2}ev_{\mathrm{F}} \bw\sim r^{-(d+2)/2}(T/T_{\mathrm{F}})^{(d+1)/2} & T_1<T<  T_4 \\
 \sqrt{\nu/2}ev_{\mathrm{F}}\sim T^{0}  & T_4 <T\ll T_{\mathrm{F}} 
\end{array} \right. .
\eeal
\ee
As $\norm{|J_x\ra}\approx\sqrt{\nu/2}e v_{\mathrm{F}}\sim T^0$, we find that when $T<T_4$ the charge current is coherent; when $T>T_4$ the charge current is incoherent. To understand this, we note that $T_4$ is the temperature above which the phonon momentum starts to dominate the total momentum, however, the charge is fully carried by the electron momentum, and such mismatch in charge current and momentum operator results in an incoherent current. Note that when charge current is coherent, the incoherent current has different scalings from $|J_{\inc}\ra\sim \ba$ to $|J_{\inc}\ra\sim \bw$ signaled by the critical temperature $T_1$. This comes from the competition between relative weights of total momentum in Fermi surface fluctuation and  phonon momentum.

Next, the incoherent heat current is given by 
\be
\beal
\norm{|Q_{\inc}\ra}&\equiv\norm{|Q_x\ra-\frac{\la p_x |Q_x\ra}{\la p_x|p_x\ra}|p_x\ra}
\approx\sqrt{\frac{\nu}{2}p_{\mathrm{F}}^2v_{\mathrm{F}}^4}\sqrt{(\ba^2+r^4\bw^2)-\frac{(\ba^2+r^2\bw^2)^2}{1+\bw^2}}\\
&\approx 
 \sqrt{\frac{\nu}{2}p_{\mathrm{F}}^2v_{\mathrm{F}}^4\ba^2}\sim T/T_{\mathrm{F}},\quad T\ll  T_{\mathrm{F}}.
\eeal
\ee
Since
\be
\norm{|Q_x\ra}=\sqrt{\frac{\nu}{2}p_{\mathrm{F}}^2v_{\mathrm{F}}^4(\ba^2+r^4\bw^2)}
\approx\left\{ \begin{array}{ll}
 \sqrt{\frac{\nu}{2}p_{\mathrm{F}}^2v_{\mathrm{F}}^4\ba^2}\sim T/T_{\mathrm{F}} & T<  T_7 \\
\sqrt{\frac{\nu}{2}p_{\mathrm{F}}^2v_{\ph}^4\bw^2}\sim r^{-(d-2)/2}(T/T_{\mathrm{F}})^{(d+1)/2} & T_7 < T\ll T_{\mathrm{F}}
 \end{array} \right. ,
\ee
we find that when $T\ll   T_7 $ the heat current is incoherent; when $T\gg T_7 $ the heat current is coherent. 
In contrast, the energy current 
\be
|J_{\mathrm{E}x}\ra=\frac{\mu}{e}|J_x\ra+|Q_x\ra.
\label{eqn:energycurrent}
\ee
is dominated by the charge current when $T\ll T_4$ (and is thus trivially coherent); moreover,  $|J_{\mathrm{E}x}\ra\approx \mu |J_x\ra $ has the same coherent-incoherent transition at $T\sim T_4$.  When $T\gg T_8$, however, $|J_{\mathrm{E}x}\ra\approx |Q_x\ra$.  Because when $T\gg T_8$, the sound velocity $v_{\mathrm{s}} \approx v_{\ph}/\sqrt{d}$ arises from a thermodynamic equation of state which is dominated by the phonons, and there is a small incoherent thermal conductivity, we claim that the fluid behaves as a nearly charge-neutral, approximately relativistic \cite{Hartnoll:2007ih} fluid.

Within the framework of kinetic theory formalism, the incoherent conductivity matrix $\Sigma_0$ can be written as
\be
\Sigma_{AB}=\left(\begin{array}{cc}
\sigma_{\inc} & T\a_{\inc} \\
T\a_{\inc} & T\bar{\kappa}_{\inc} \end{array}\right)=
\left(\begin{array}{c}
\la J_{\inc}| \\
\la Q_{\inc}| \end{array}\right)W_{\text{e-ph}}^{\prime -1}
\left(\begin{array}{cc}
|J_{\inc}\ra & |Q_{\inc}\ra \end{array}\right),
\label{eqn:incmat}
\ee
where all the quantities point to $x$ direction. Explicit expressions and scalings of each element are: 
\begin{subequations}
\begin{align}
\sigma_{\inc}&
\approx \frac{e^2v_{\mathrm{F}}^2\nu}{(1+\ba^2+\bw^2)^2} \left\{\left[(1-p_{\mathrm{F}}\pa_p\ln v_{\mathrm{F}})\ba^2+\bw^2\right]^2\ga^{-1}_{\e}+\left[1-(1+\bw^2)p_{\mathrm{F}}\pa_p\ln v_{\mathrm{F}}\right]^2\ba^2\gamma^{\prime -1}_{\e}+\bw^2\ga^{-1}_{\ph}\right\}\\
&\approx \left\{ \begin{array}{ll}
e^2(v_{\mathrm{F}}-p_{\mathrm{F}}\pa_pv_F)^2\nu \ga^{-1} r^{-2}\ba^4 \sim r^{-2}(T/T_{\mathrm{F}})^{-(d-2)} & T<T_1/r \\
 2e^2v_{\mathrm{F}}^2\nu \ga^{-1} \bw^4 \sim r^{-2(d+2)}(T/T_{\mathrm{F}})^{d} &  T_1/r<T<T_4 \\
  2e^2v_{\mathrm{F}}^2\nu \ga^{-1} \sim (T/T_{\mathrm{F}})^{-(d+2)} & T_4<T<T_{\text{BG}}\\
  2e^2v_{\mathrm{F}}^2\nu \bar{\gamma}^{-1}\sim (T/T_{\mathrm{F}})^{-1} & T_{\text{BG}}<T\ll  T_{\mathrm{F}}
\end{array} \right. , \nonumber\\[2ex]
T\bar{\kappa}_{\inc}&
\approx \frac{\nu p_{\mathrm{F}}^2v_{\mathrm{F}}^4}{(1+\ba^2+\bw^2)^2}\left\{ (\ba^2+r^2\bw^2)^2\ga_{\e}^{-1}+ (1+\bw^2)^2\ba^2\ga_{\e}'^{-1}+(r^2-\ba^2)^2\bw^2\ga_{\ph}^{-1} \right\}  \\
&\approx \left\{ \begin{array}{ll}
 \nu v_{\mathrm{F}}^4p_{\mathrm{F}}^2\ga^{-1}\frac{\ba^4}{r^2} \sim r^{-2}(T/T_{\mathrm{F}})^{-(d-2)}  & T<T_{\text{BG}} \\
  \nu v_{\mathrm{F}}^4p_{\mathrm{F}}^2\bar{\gamma}^{-1} \ba^2\sim T/T_{\mathrm{F}}  & T_{\text{BG}}<T\ll  T_{\mathrm{F}}
\end{array} \right. ,\nonumber\\[2ex]
T\a_{\inc}& \approx -\frac{e\nu v_{\mathrm{F}}^3p_{\mathrm{F}}}{(1+\ba^2+\bw^2)^2}\Big\{\left[(1-p_{\mathrm{F}}\pa_p\ln v_{\mathrm{F}})\ba^2+\bw^2\right](\ba^2+r^2\bw^2)\ga_{\e}^{-1}\\
&\qquad +\left[1-(1+\bw^2)p_{\mathrm{F}}\pa_p\ln v_{\mathrm{F}}\right](1+\bw^2)\ba^2\ga_{\e}'^{-1}+(r^2-\ba^2)\bw^2\ga_{\ph}^{-1} \Big\} \nonumber\\
&\approx \left\{ \begin{array}{ll}
-e\nu v_{\mathrm{F}}^2p_{\mathrm{F}}(v_{\mathrm{F}}-p_{\mathrm{F}}\pa_p v_{\mathrm{F}}) \ga^{-1}r^{-2}\ba^4\sim r^{-2}(T/T_{\mathrm{F}})^{-(d-2)} & T<T_3\\
-2e\nu v_{\mathrm{F}}^3p_{\mathrm{F}} \ga^{-1}r^2\bw^4 \sim r^{-2(d+1)}(T/T_{\mathrm{F}})^d  & T_3<T<T_4 \\
-2e\nu v_{\mathrm{F}}^3p_{\mathrm{F}} \ga^{-1}r^2 \sim r^{2}(T/T_{\mathrm{F}})^{-(d+2)}  & T_4<T<T_{\text{BG}} \\
e\nu v_{\mathrm{F}}^3p_{\mathrm{F}} \bar{\gamma}^{-1}\ba^2 \sim T/T_{\mathrm{F}} & T_{\text{BG}}<T\ll T_{\mathrm{F}}\\
\end{array} \right. . \nonumber
\end{align}
\end{subequations}
We observe that one might naively expect that for a Galilean dispersion relation in which $v_{\mathrm{F}} -  p_{\mathrm{F}}\pa_p v_{\mathrm{F}} = 0$, that both $\sigma_{\mathrm{inc}}$ and $\alpha_{\mathrm{inc}}$ must vanish identically since there is no incoherent current:  electrical current and momentum are proportional \cite{hartnoll2018holographic}.  However, this is not true, because the presence of phonons dispersion relation necessarily leads to breaking of Galilean invariance. Notice that naive estimation like $\sigma_{\mathrm{inc}}\sim \gamma^{-1}\norm{|J_x\ra}^2$ suitable for Fermi liquid is no longer true for electron-phonon fluids because of the multiple scattering rates. 

Now we are ready to compute the diffusion constant
\be
\Gamma=\frac{2\frac{d-1}{d}\eta+\zeta}{M}+\frac{\Sigma}{M},
\label{eqn:diffusion}
\ee
where
\be
\Sigma\equiv\frac{\rho_A\chi^{-1}_{AC}\Sigma_{CD}\chi^{-1}_{DB}\rho_B}{v_{\mathrm{s}}^2}
\ee 
is a contribution to the decay rate arising from incoherent conductivities.
After some algebra, we find that
\be
\Gamma\sim\left\{ \begin{array}{ll}
\ga^{-1} \sim (T/T_{\mathrm{F}})^{-(d+2)}  & T<T_{\text{BG}}\\
  \bar{\gamma}^{-1} \sim (T/T_{\mathrm{F}})^{-1}&  T_{\text{BG}}<T<T_6\\
 \bar{\gamma}^{-1}r^{-2}\bw^{-2} \sim r^d (T/T_{\mathrm{F}})^{-(d+2)}   & T_6<T\ll T_{\mathrm{F}}\\
\end{array} \right. .
\ee
The scaling is shown in \figref{fig:hydro}(c). When $T<T_4$, the shear viscosity, given by the electron modes, controls the diffusion of the fluid, i.e. $\eta > \Sigma$. When $T>T_4$, the diffusion constant starts to be dominated by the incoherent conductivity, i.e. $\eta< \Sigma$, but the scaling do not change. The interplay between shear viscosity and incoherent conductivity implies a transition from coherent to incoherent, giving rise to a breakdown of Galilean invariance.

%

\subsection{Plasmons}\label{sec:plasmons}
Although short-range interactions between individual electrons are assumed to be negligible in our model (for pedagogical purposes), the long-range Coulomb interaction, responsible for the density-density interaction gives rise to qualitative changes in the hydrodynamic dispersion relations, and so we will briefly address what happens in the presence of unscreened and long-range Coulomb interactions. We account for the long-range Coulomb interaction by replacing the chemical potential with the external electrochemical potential \cite{PhysRevB.93.245153,PhysRevB.93.075426,PhysRevB.78.115419} 
\be
\pa_i \mu \to \pa_i\mu-F_{\text{ext}}
\ee
where 
\be
F_{\text{ext}}=-\pa_i \int \ud^d y \frac{e^2}{|x-y|}(\rho(y)-\rho_0) .
\ee
Fourier transforming \eqnref{visc}, we obtain the frequency-momentum space hydrodynamic equation 
\begin{subequations}\label{eqn:plasmon}
\begin{align}
-\i \omega \chi_{AB}\d \mu_B+\i k \rho_A \d u_i +k^2 \tilde{\Sigma}_{AB}\d\mu_B&=0\\
-\i \omega M \d u_i +\i k \tilde{\rho}_A\d \mu_A+ \left(\eta  + \frac{2d-2}{d}\zeta\right)k^2 \d u_i&=0
\end{align}
\end{subequations}
where we assume $\b{k}$ is parallel to  $\d \b{u}$ and 
\be
\tilde{\rho}_1=(1+U(k)\nu)\rho_1,\quad  \tilde{\rho}_2=\rho_2,\quad \tilde{\Sigma}_{A1}=(1+U(k)\nu)\Sigma_{A1},\quad \tilde{\Sigma}_{A2}=\Sigma_{A2},
\ee
with $F_{\text{ext}}(k)=-\i k U(k)\nu \d \mu$. Note that for tilded operators, only components related to the chemical potential are corrected by the long-range Coulomb interaction. We take $d=2$ for illustration. For small wave vector, we can approximate 
\be
(1+U(k)\nu)k^2\approx 2\pi e^2 \nu |k|.
\ee
\begin{figure}[t]
 \includegraphics[width=.4\linewidth]{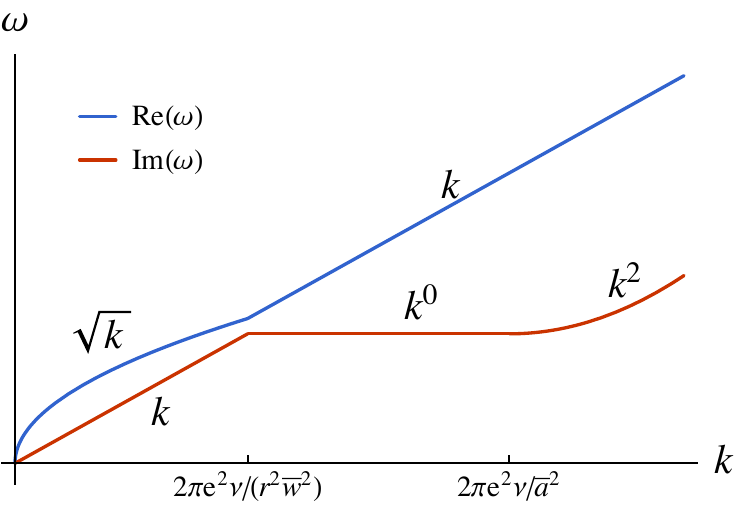}
\caption{Plasmon dispersion relation for electron-phonon fluid in the regime $T_6<T\ll T_{\mathrm{F}}$. }
\label{fig:plasmon}
\end{figure}
Solving \eqnref{eqn:plasmon}, we have the dispersion relation of plasmons
\be
\beal
\omega_{\text{plasmon}}&=\pm\sqrt{\frac{\tilde{\rho}_A\chi^{-1}_{AB}\rho_B}{M}k^2}-\i \left(\frac{2(d-1)\eta}{dM}+\frac{\tilde{\rho}_A\chi^{-1}_{AC}\tilde{\Sigma}_{CD}\chi^{-1}_{DB}\rho_B}{\tilde{\rho}_A\chi^{-1}_{AB}\rho_B}\right)k^2\\
&\approx \left\{ \begin{array}{ll}
\displaystyle \pm \frac{v_{\mathrm{F}}}{\sqrt{d}}\sqrt{2\pi e^2\nu |k|}-\i \nu\mu^2\ga^{-1}(k^2+\ba^4(2\pi e^2\nu)|k|) & T<T_1 \\
\displaystyle \pm \frac{v_{\mathrm{F}}}{\sqrt{d}}\sqrt{2\pi e^2\nu |k|}-\i \nu\mu^2\ga^{-1} (k^2+\bw^4(2\pi e^2\nu)|k|) & T_1<T<T_4 \\
\displaystyle \pm \frac{v_{\mathrm{F}}}{\bw\sqrt{d}}\sqrt{2\pi e^2\nu |k|}-\i \nu\mu^2\ga^{-1}(k^2/\bw^2 +(2\pi e^2\nu) |k|) & T_4<T<T_{\text{BG}} \\
\displaystyle \pm \frac{v_{\mathrm{F}}}{\bw\sqrt{d}}\sqrt{2\pi e^2\nu |k|}-\i \nu\mu^2\bar{\gamma}^{-1}(k^2/\bw^2+ (2\pi e^2\nu) |k|) & T_{\text{BG}}<T<T_6 \\
\displaystyle \pm \frac{v_{\mathrm{F}}}{\bw\sqrt{d}}\sqrt{2\pi e^2\nu |k|+r^2\bw^2k^2}-\dfrac{\i \nu\mu^2}{\bar{\gamma}} \dfrac{\ba^4 k^2 + (2\pi e^2\nu)^2}{2\pi e^2 \nu + r^2\bw^2|k|}|k| 
& T_6<T\ll T_{\mathrm{F}} \\
\end{array} \right. .
\eeal
\ee
In this expression we have used the fact that $\eta \gg \zeta$ to simplify the result.

For $T < T_6$, $\omega_{\text{plasmon}}\sim \sqrt{k}$ manifests the conventional plasmon dispersion relation \cite{ando,hwang1,hwang2}
\begin{equation}
    \omega_{\mathrm{plasmon}} = \sqrt{\frac{2\pi e^2 n}{m} |k|} + \cdots;
\end{equation}
note we have used the identity (\ref{eq:nuidentity}).  Importantly, however, the prefactor of $|k|$ is also affected by the large contribution of phonons to $M$ (and therefore the effective value of $m$) above $T>T_4$, and so it will exhibit the same dramatic $T$-dependence as the sound velocity \eqnref{eqn:soundv}.  This provides a clear difference with the usual electron fluid where electron-electron interactions dominate \cite{PhysRevB.93.245153,phan2013ballistic}. When the temperature is below $ T_4$, the hydrodynamic is coherent, and the imaginary part, known as the plasmon decay, is the usual sound wave decay \eqnref{eqn:diffusion}, but with a crossover to the more severe plasmon decay $\d \omega_{\text{plasmon}}\sim -\i \ga^{-1}|k|$ at ultra low $k$. However, when $T>T_4$, this plasmon decay starts to dominate indicating the increasingly incoherent nature of the charge current.

As the temperature keeps increasing, the plasmon mode starts to transform into the phonon-limited sound mode $\omega\sim v_{\ph}/\sqrt{d}$ when $T>T_6$,  at short enough distance scales.   The long distance physics with small enough $k$ ($r^2\bw^2|k|<2\pi e^2\nu$) is always controlled by the long-range Coulomb interactions. Interestingly, the plasmon decay in the high temperature regime shows a dramatically different behavior. In the regime $T_6<T\ll T_{\mathrm{F}}$, the second term $\ba^4$ in imaginary part could be neglected resulting in a $k$-independent plasmon decay in $d=2$ at short distance. Similar behavior exists in hydrodynamics with momentum relaxed by impurities \cite{hartnoll2018holographic}, where such $k$-independent decay is given by $-\i\tau_{\text{imp}}^{-1}$. However, here, it is induced purely through the long-range Coulomb interaction. Notice that such $k$-independent plasmon decay will crossover to the incoherent plasmon decay $\sim - \i|k|$ at long distance (same as the crossover in the real part), while crossover to the sound wave decay $\sim - \i k^2$ at short distance ($\ba^2|k|>2\pi e^2\nu$). The behaviors are summarized in \figref{fig:plasmon}. 

In three dimensions ($d=3$), the generalization of the above computation leads to:
\be
\omega_{\text{plasmon}}
\approx \left\{ \begin{array}{ll}
\displaystyle \pm \frac{v_{\mathrm{F}}}{\sqrt{d}}\sqrt{4\pi e^2\nu}-\i \nu\mu^2\ga^{-1}(k^2+\ba^4(4\pi e^2\nu)) & T<T_1 \\
\displaystyle \pm \frac{v_{\mathrm{F}}}{\sqrt{d}}\sqrt{4\pi e^2\nu}-\i \nu\mu^2\ga^{-1} (k^2+\bw^4(4\pi e^2\nu)) & T_1<T<T_4 \\
\displaystyle \pm \frac{v_{\mathrm{F}}}{\bw\sqrt{d}}\sqrt{4\pi e^2\nu}-\i \nu\mu^2\ga^{-1}(k^2/\bw^2 +(4\pi e^2\nu)) & T_4<T<T_{\text{BG}} \\
\displaystyle \pm \frac{v_{\mathrm{F}}}{\bw\sqrt{d}}\sqrt{4\pi e^2\nu}-\i \nu\mu^2\bar{\gamma}^{-1}(k^2/\bw^2+ (4\pi e^2\nu)) & T_{\text{BG}}<T<T_6 \\
\displaystyle \pm \frac{v_{\mathrm{F}}}{\bw\sqrt{d}}\sqrt{4\pi e^2\nu+r^2\bw^2k^2}-\dfrac{\i \nu\mu^2}{\bar{\gamma}} \dfrac{\ba^4 k^4 + (4\pi e^2\nu)^2}{4\pi e^2 \nu + r^2\bw^2 k^2} 
& T_6<T\ll T_{\mathrm{F}} 
\end{array} \right. .
\ee
The plasmon dispersion relation becomes $k$-independent, and due to the incoherent conductivity, it also picks up a $k$-independent finite lifetime (above $T_1$). This might explain the rapid plasmon decay in recent experiments on 3d strongly-correlated electron systems \cite{abbamonte19,husain2020coexisting}, if the electron-phonon interaction is not negligible.

We emphasize that the imaginary part of the plasmon dispersion relation is also affected by impurities, recombination processes, etc., beyond our kinetic theory treatment.  So it may be much easier to look for the unconventional real dispersion relation modification that we predict above in a near-term experiment.

\section{Thermoelectric transport}\label{sec:WF}
In this section, we carefully address the interplay of both the electron- and phonon-impurity scattering and  the electron-phonon scattering, but neglect the electron-electron scattering, as we have previously.   We first summarize our results, and then compare to existing experimental data and compare and contrast our results from transport with our predictions in Section \ref{sec:hydro} for other types of experiment.

\subsection{Formal results}
We remind the reader that the vector without tilde is the orthonormal basis after Gram-Schmidt method (see \eqnref{eqn:norm}), and obtain
\bes
\be
\la 0|W^\e_{\text{imp}}|0\ra_{\e}=\Gamma_{\e},
\ee
\be
\la 1|W^\e_{\text{imp}}|1\ra_{\e}=\Gamma_{\e},
\ee
\be
\la 0|W^\e_{\text{imp}}|1\ra_{\e}=\frac{\pi T}{\sqrt{3}}(\frac{\pa\Gamma_{\e}}{\pa \mu}-\frac{\pa_pv_{\mathrm{F}}}{v_{\mathrm{F}}^2}\Gamma_{\e})\equiv b,
\ee
\be
\la 1|W^{\ph}_{\text{imp}}|1\ra_{\ph}=\Gamma_{\ph},
\ee
\ees
where $\Gamma_{\e,\ph}$ is the impurity scattering rate based upon relaxation time approximation. We work in a simple limit
\begin{equation}
    \Gamma_{\text{e,ph}}\ll \gamma_0,
\end{equation}
according to \eqnref{eqn:gamma}, such that the electron-phonon scattering rate could be greater than the impurity scattering rate at an appropriately temperature.
Recalling \eqnref{eqn:Weph},
the total collision integral is given by
\be
W=W^\e_{\text{imp}}+W^{\ph}_{\text{imp}}+W_{\text{e-ph}}',
\ee
where $W^\prime_{\text{e-ph}}$ denotes the momentum-conserving collision integral studied before.  The thermoelectric conductivity matrix can be written as \cite{PhysRevB.97.245128}
\be
\left(\begin{array}{cc}
\sigma_{xx} & T\a_{xx} \\
T\a_{xx} & T\bar{\kappa}_{xx} \end{array}\right)=
\left(\begin{array}{c}
\la J_x| \\
\la Q_x| \end{array}\right)W^{-1}
\left(\begin{array}{cc}
|J_x\ra & |Q_x\ra \end{array}\right).
\label{eq:mat}
\ee
The detailed expression can be found in \appref{app:wf}. Focusing on $T<T_{\text{BG}}$ for simplicity, we observe that
\begin{subequations}\label{eqn:conductivity}
\begin{align}
    \sigma_{xx}&=e^2\frac{\nu}{2}v_{\mathrm{F}}^2\frac{\Gamma_{\ph}+\ga/\bw^2}{\Gamma_{\e}(\Gamma_{\ph}+\ga/\bw^2)+\Gamma_{\ph}\ga}+\mO(T^2),\\
    T\bar{\kappa}_{xx}&=\frac{\nu}{2}p_{\mathrm{F}}^2v_{\mathrm{F}}^4\left(\frac{\ba^2}{\Gamma_{\e}+\ga r^2/\ba^2}+\frac{r^4\bw^2(\Gamma_{\e}+\ga)}{\Gamma_{\e}(\Gamma_{\ph}+\ga/\bw^2)+\Gamma_{\ph}\ga}+2r^2\ba\frac{-(b-\gamma r^2/\ba)\gamma+(\Gamma_{\e}+\gamma)\gamma r^2/\ba}{(\Gamma_{\e}+\ga r^2/\ba^2)[\Gamma_{\e}(\Gamma_{\ph}+\ga/\bw^2)+\Gamma_{\ph}\ga]} \right),\label{eqn:barkappa} \\ 
    T\a_{xx}&=-e\frac{\nu}{2}p_{\mathrm{F}}^2v_{\mathrm{F}}^2\left(\frac{r^2\ga}{\Gamma_{\e}(\Gamma_{\ph}+\ga/\bw^2)+\Gamma_{\ph}\ga}+\frac{\ba^2}{\Gamma_{\e}+\ga r^2/\ba^2} +\ba\frac{-(\Gamma_{\ph}+\ga/\bw^2)(b-\ga r^2/\ba)+\ga^2 r^2/(\ba\bw^2)}{(\Gamma_{\e}+\ga r^2/\ba^2)[\Gamma_{\e}(\Gamma_{\ph}+\ga/\bw^2)+\Gamma_{\ph}\ga]}\right).
\end{align}
\end{subequations}
Several remarks follow the above equations. First, unlike the electron-electron interaction \cite{PhysRevB.97.245128}, the electrical conductivity can be more strongly affected by the electron-phonon interaction if $\Gamma_{\mathrm{ph}}$ is not negligible.  More drastic is the correction to thermal conductivity, as well as thermoelectric conductivity. Second, all the thermodynamic properties have a divergence in the limit $\Gamma_{\text{e,ph}}\to 0$. This is a universal result for any hydrodynamic fluid with translation symmetry \cite{hartnoll2018holographic}. However, for the experimental thermal conductivity
\begin{equation}
    T\kappa_{xx}=T\bar{\kappa}_{xx}-\frac{(T\a_{xx})^2}{\sigma_{xx}}\approx \frac{\nu}{2}p_{\mathrm{F}}^2v_{\mathrm{F}}^4\left(\frac{\ba^2}{\Gamma_{\e}+\ga r^2/\ba^2}+\frac{r^4\bw^2}{\Gamma_{\ph}+\ga/\bw^2} \right),
\end{equation}
such divergence disappears consistently. Consequently, the ratio of the experimental thermal and electrical conductivity, i.e. the Lorenz number $L$, approaches zero in the (hydrodynamic) limit $\Gamma_{\mathrm{e,ph}}\to 0$. In particular, a controlled way to estimate how clean a material should be to have a divergent Lorenz number is to compare the non-diverging and diverging term inside the thermal conductivity \eqnref{eqn:barkappa}. By assuming $\Gamma_{\ph}=0$ in the first place, we find that as long as
\begin{equation}
    \Gamma_{\e}\lesssim   \frac{\gamma r^4 \bw^2}{\ba^2}
    ,\quad T<T_{\mathrm{BG}}
\end{equation}
the Lorenz number $L\ll L_0$, where $L_0=\pi^2/3e^2$ is the non-interacting Lorenz number, and can become arbitrarily small. This effect is well understood \cite{hartnoll2018holographic}: as $\Gamma_{\mathrm{e}}\rightarrow 0$, at any finite charge density $\sigma_{xx}$ is divergent while $\kappa_{xx}$ remains finite.  
Another feature that our model hold is that in the non-interacting limit $\ga\to 0$, which corresponds to $T\to 0$, both the WF law and Mott law are simultaneously recovered so long as the impurity scattering rate is not zero, as in a normal metal  \cite{PhysRevB.97.245128}.
\begin{figure}[t]
 \includegraphics[width=1.\linewidth]{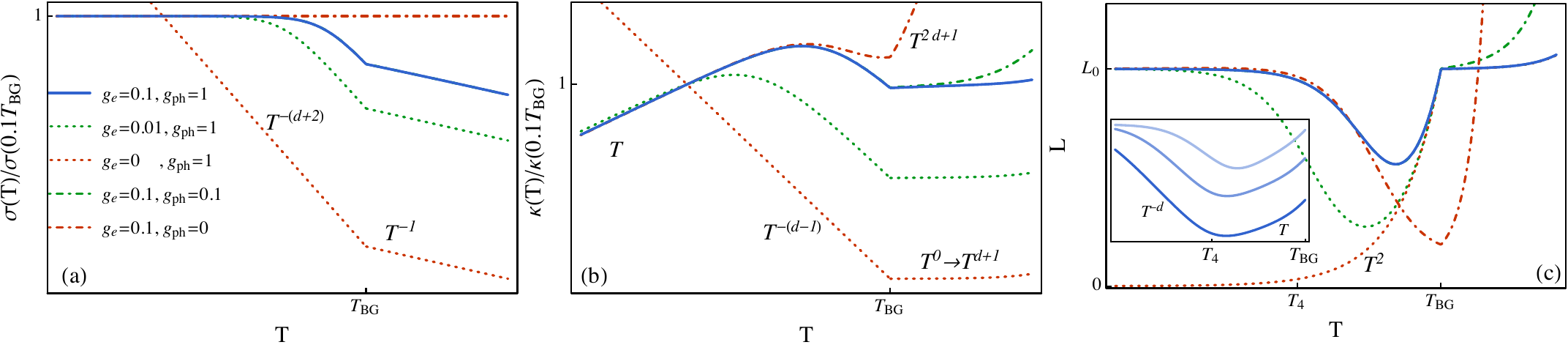}
\caption{Thermoelectric conductivity versus temperature with five different impurity scattering rates. (a) Log-log plot of electrical conductivity against temperature. The dot-dashed green line overlaps with the blue line. (b) Log-log plot of open-circuit thermal conductivity against temperature. (c) Log-linear plot of Lorenz number versus temperature. The inset (log-log plot) shows that the dip is approaching $T_4$ as lowering from $g_\e=g_\ph=10^{-3}$, $10^{-4}$ to $10^{-5}$ (from top to bottom). The kink at $T_{\mathrm{BG}}$ comes from the fact that the electron-phonon scattering rate in our model has a jump there. We take $r=10^{-3}$ for the plots.}
\label{fig:lorenz2}
\end{figure}
We determine the Lorenz number with finite impurity scattering: 
\begin{equation}\label{eqn:lorenz}
L\equiv\frac{\kappa_{xx}}{T\sigma_{xx}}\approx \left\{ \begin{array}{ll}
 \displaystyle L_0\left( \frac{\Ge(\Gph+\ga/\bw^2)+\Gph\ga}{(\Ge+\ga r^2/\ba^2)(\Gph+\ga/\bw^2)}+\frac{r^4\bw^2}{\ba^2}\frac{\Ge(\Gph+\ga/\bw^2)+\Gph\ga}{(\Gph+\ga/\bw^2)^2} \right)& ~ T<T_{\text{BG}}\\
  \displaystyle L_0\left( \frac{\Ge(\Gph+\bar{\ga}/\bw^2)+\Gph\bar{\ga}}{(\Ge+\bar{\ga} )(\Gph+\bar{\ga}/\bw^2)}+\frac{r^4\bw^2}{\ba^2}\frac{\Ge(\Gph+\bar{\ga}/\bw^2)+\Gph\bar{\ga}}{(\Gph+\bar{\ga}/\bw^2)^2} \right) & ~ T_{\text{BG}}<T\ll T_{\mathrm{F}}
\end{array} \right. .
\end{equation}
Defining the ratios
\begin{subequations}
\begin{align}
    g_\e&\equiv\frac{\Gamma_{\e}}{r \gamma_0},\\
    g_\ph&\equiv\frac{\Gamma_{\ph}}{r \gamma_0},
\end{align}
\end{subequations}
(recall the definition of $\gamma_0$ in \eqnref{eqn:gamma0}) we sketch out the electrical and thermal conductivity, and the Lorenz number against the temperature with various $g_{\e,\ph}$ in \figref{fig:lorenz2}.

\subsection{Comparison to canonical transport theory}
Let us now compare our predictions to the standard theory of thermoelectric transport in metals with electron-phonon scattering.  The qualitative shape of these curves is extremly similar to conventional metals \cite{ziman2001electrons}, as we now explain.  Firstly, at low temperatures, assuming $g_{\mathrm{e}} \ne 0$, we see that \begin{equation}
    \sigma_{xx} \approx \frac{\nu e^2 v_{\mathrm{F}}^2}{2\Gamma_{\mathrm{e}}},
\end{equation}
and $L\approx L_0$ -- this is because at low temperatures, electron-impurity scattering controls transport.  The Wiedemann-Franz law holds, in accordance with experiments.

At intermediate temperatures approaching (but below) $T_{\mathrm{BG}}$, we see a dip in $L/L_0$ below 1.  One of the primary reasons for this is that the thermal conductivity obtains a correction due to energy-relaxing scattering events that occur with rate $T^3$ (in $d=3$ dimensions).   This leads to $\bar\kappa_{xx} \sim T^{1-d}$,  which is the standard $\bar\kappa_{xx} \sim T^{-2}$ scaling that arises at intermediate temperatures in a standard metal.  We identify the temperature where $L$ starts to deviate from $L_0$ as
\begin{equation}
    T^{*}\approx  g_\e^{1/d}T_{\mathrm{BG}}. 
\label{eq:Tstar}
\end{equation}
This comes from $\Gamma_\e\approx \gamma r^2/\ba^2$ and should be compared with $T_i$ in \cite{PhysRevB.99.085104}.
Meanwhile, the minimum value of $L$ is located at $T_{\mathrm{min}}=xT_{\mathrm{BG}}$ where $x$ is the solution of 
\begin{equation}
    -d g_\e+(d+2)g_{\e}x^2+2x^{d+2}=0.
\label{eq:Tmin}
\end{equation} 
Both $T^{*}$ and $T_{\mathrm{min}}$ are not universal and depends on the impurity strength; they vanish in the clean limit $g_\e\to 0$, where $L\approx L_0 \ba^2/r^2$ at low temperature scales as $\sim T^2$. On the other hand, lowering $g_{\mathrm{ph}}$ hardly affects $L$ below $T_{\mathrm{BG}}$, but makes $L$ increase more rapidly above $T_{\mathrm{BG}}$. When $T>T_{\mathrm{BG}}$, $L$ will eventually surpass $L_0$ due to the plethora of phonon excitations.

In contrast, if we set $g_{\mathrm{e}}=0$, the electrical conductivity $\sigma$ in the clean limit $g_{\mathrm{e}}=0$ scales as $\sim T^{-d-2}$ below $T_{\mathrm{BG}}$ while $\sim T^{-1}$ above $T_{\mathrm{BG}}$.  The reason for this is essentially the conventional one given in the literature: the phonons relax away momentum, and the small-angle scattering of an electron by phonons dominates electrical transport.  In contrast, for $T<T_{\mathrm{BG}}$, the thermal conductivity $\kappa \sim T^{-d+1}$, for the same reason described above.   Similar plots can be found in \cite{PhysRevB.99.085104}. 

One interesting result that is absent in the literature is shown in the inset of \figref{fig:lorenz2}(c). Instead of tuning  $g_\e$ down alone, we gradually decrease $g_\e\sim g_\ph$ together, and we find that the minimum of $L$ approaches $T_4$. Such behavior and the scaling of $L$ near $T_4$ can been seen from the asymptotic expression
\begin{equation}
    L\approx \left\{ \begin{array}{ll}
 \displaystyle L_0 \frac{1}{\gamma r^2 /\ba^2}\sim T^{-d}& ~ T\lesssim T_{4}\\
  \displaystyle L_0 \frac{\bw^2}{\gamma r^2 /\ba^2} \sim T& ~ T\gtrsim T_4
\end{array} \right. ,
\end{equation}
where we assumed $\Gamma_\ph\sim\Gamma_\e\ll \gamma$. The change in $T$ scaling arises when $\bw \gtrsim 1$ at $T=T_4$, and so this physics arises from the same mechanism which gives rise to the non-trivial $T$-dependence of the sound speed discussed in \secref{sec:hydro}. Hence, whenever the electron and phonon momentum relaxing scattering rates are comparable and small compared to momentum-conserving scattering rates, we can unambiguously state that $T_4$ instead plays the role as a \textit{universal} temperature for the strong violation of WF law. 


One difference between our work and \cite{PhysRevB.99.085104} is what happens at high $T$.  In \cite{PhysRevB.99.085104}, they argue that $L/L_0$ decreases due to thermal smearing of the Fermi surface.  This effect lies at temperatures $T\sim T_{\mathrm{F}}$ which are not studied in this work.  Without electron-phonon umklapp scattering, $L/L_0$ will eventually exceed 1; this arises when the phonons dominate the energy density of the hybrid fluid, and the fluid becomes a nearly charge-neutral relativistic fluid.  The hydrodynamic equations become similar to those describing the Dirac fluid in charge neutral graphene \cite{Crossno_2016}.   

A transport regimes that differs from the canonical expectations, but which we expect is not likely to realize in experiment, can be found in Appendix \ref{app:newregime}.

Overall, due to the similarity between our theory and the conventional theory of transport in a metal with electron-phonon interactions, we emphasize that extracting hydrodynamic signatures of an electron-phonon fluid from conventional transport measurements is delicate.  Nevertheless, as we will discuss below, our theory above can still be used to analyze experimental data in relevant materials, and we will argue that many materials lie in the very interesting regime $T_4<T<T_{\mathrm{BG}}$.

\subsection{Comparison to real materials}


Having discussed the key predictions of our theory of homogeneous transport, let us now comment on experiments in a number of materials where evidence for electron-phonon hydrodynamics has been obtained.  Our main conclusion is that it is likely that at least $\mathrm{PtSn}_4$ and $\mathrm{WP}_2$ are hydrodynamic at temperatures $T_4\lesssim T \lesssim T_{\mathrm{BG}}$.  

There are two caveats, common to both materials, that we state upfront.  (\emph{1})  Because $T_4$ and $T_{\mathrm{BG}}$ are not orders of magnitude apart, one should not expect parametric scaling regimes like those of Section \ref{sec:hydro} to be easily detected. One might hope instead to see ``kinks" in the $T$-dependence of various quantities whose relative signs/amplitudes are consistent with Section \ref{sec:hydro}.  (\emph{2}) These materials are highly anisotropic.  We, like other authors, are attempting to use an isotropic theory to approximately describe them;  there is no quantitative guarantee that anisotropy leads to only negligible corrections.

Having stated our main conclusions and their limitations, we first turn to $\mathrm{PtSn}_4$, which was studied in \cite{fu2018thermoelectric}.  Above $T\approx 25$ K, the resistivity scales as $\rho\sim T$, manifesting the linear-in-T behavior expected at $T>T_{\mathrm{BG}}$ \cite{PhysRevB.99.085105}. Since the thermal conductivity scaling $\kappa\sim T^{-2}$ also ends at $T\approx 25$ K (recall \figref{fig:lorenz2}(b)), we conclude that  $T_{\mathrm{BG}}\approx 25$K. Below $T\approx 8$ K, the resistivity saturates to the residual resistance, and the thermal conductivity scales as $\kappa\sim T$. The WF law also remains valid due to the dominant impurity scattering below $T\approx 8$ K. Thus the relative electron impurity strength can be extracted from this temperature turning point by noticing that the maximum point of $\kappa$ is given by $T\approx  g_e^{1/d}T_{\mathrm{BG}}$, the same as \eqnref{eq:Tstar} where the Lorenz number starts to go below $L_0$; we have $g_e\approx 0.04$. Then, \eqnref{eq:Tmin} is solved to give the temperature of the minimum Lorenz number at $T_{\mathrm{min}}\approx 12.7$K.  Interestingly, this minimum is at 14 K in experiment  \cite{fu2018thermoelectric}, which is quite close to our predictions (which did not account for the complication of the Fermi surface geometry!). A careful study about the influence of Fermi surface geometry will be a future work.


Now, let us see which temperature regime (\tabref{tab:sum}) such transport phenomena would correspond to.  Using the following experimental data \cite{fu2018thermoelectric}: effective mass $m_{\mathrm{eff}} \approx 0.2 m_{\mathrm{e}}$, electron density $n\sim 4\times 10^{20} \; \mathrm{cm}^{-3}$, and Fermi velocity $v_{\mathrm{F}} \sim 4\times 10^5$ m/s, we estimated that $T_{\mathrm{F}}$ is around 5000 to 10000 K.  Hence we estimate $r=T_{\mathrm{BG}}/T_{\mathrm{F}}\sim (2.5-5)\times 10^{-3}$, thus we have $T_4= r^{1/4}T_{\mathrm{BG}}\approx 5-6$ K, and the violation of WF law in $\mathrm{PtSn}_4$ is  cleanly located within $T_4<T<T_{\mathrm{BG}}$. This is expected from various hydrodynamic calculations in \secref{sec:hydro} (see also \figref{fig:hydro}) where electrons and phonons flow as a single unified fluid. In general, we argue that as long as the impurity strength satisfies $r^{d/(d+1)}< g_{\mathrm{e}}$, the ``strong hybrid'' electron-phonon fluid would be identified in the temperature regime $T_4<T<T_{\mathrm{BG}}$. 


Next, we turn to $\mathrm{WP}_2$: first we discuss data reported in \cite{Jaoui2018,Gooth_2018}.  Taking $\rho_0\approx 5\times 10^{-9} \mathrm{\Omega\cdot cm}$, $A_5\approx 4 \times 10^{-15}\mathrm{\Omega\cdot cm \cdot K^{-5}}$ and $B_3\approx 6.5 \times 10^{-12}\mathrm{\Omega\cdot cm \cdot K^{-3}}$\cite{Jaoui2018}, and $T_{\mathrm{F}}=E_{\mathrm{F}}/k_B\approx 6.8\times 10^4$K\cite{Gooth_2018}, we obtain
\begin{subequations}
\begin{align}
    g_\e r^3&=\frac{\rho_0}{B_3T_{\mathrm{F}}^3}\approx 2.4 \times 10^{-12},\\
    g_\e r^5&=\frac{\rho_0}{A_5T_{\mathrm{F}}^5}\approx 8.6 \times 10^{-19}.
\end{align}
\end{subequations}
Thus $r\approx 6\times 10^{-4}$ and $g_\e \approx 0.01$. Moreover, $T_{\mathrm{BG}}=rT_{\mathrm{F}}\approx 40$K, and $T_{\mathrm{min}}\approx 16$K by solving \eqnref{eq:Tmin}.  Note that our estimate of 40 K for $T_{\mathrm{BG}}$ is compatible with the ab initio estimate for a Debye temperature in this material \cite{coulter}, which plays qualitatively the sam role of modifying the $T$-dependence in scattering mechanisms. We see that the temperature for the minimum point of Lorenz number well matches the experimental results of \cite{Gooth_2018}. Hence, it is plausible for the transport in $\mathrm{WP_2}$ to exist within electron-phonon hydrodynamics, especially when the violation of WF law happens above $T_4\approx 6$K.

Another recent experiment found sommewhat similar results \cite{Jaoui2018}, yet gave a quite different interpretation, suggesting that electron-electron scattering is responsible for the deviation from the WF law in $\mathrm{WP_2}$.  Firstly, \cite{Jaoui2018} observes that at low temperatures, there is a mismatch between $T^2$ prefactors in electrical and thermal resistivity, suggesting that momentum-conserving electron-electron scattering is important.  This is not in conflict with the ab initio proposal that phonons dominate at higher temperatures \cite{coulter}.  Secondly, \cite{Jaoui2018} argues that the ratio of $B_3$ and $A_5$ is similar in $\mathrm{WP}_2$ and Ag, while the latter does not have a significant deviation in $L/L_0$ at low $T$.  However, we emphasize that in our transport theory, the coefficients $A_3$ and $B_5$ are robust and not sensitive to the hydrodynamic regime; rather, it is the small values of $\Gamma_{\mathrm{e}}$ and $\Gamma_{\mathrm{ph}}$ that induce electron-phonon hydrodynamics.  

Both experiments \cite{Gooth_2018,Jaoui2018} agree on the scale at which $L/L_0$ is minimal.  If electron-phonon hydrodynamics does arise in this material, it is likely in the range $T_4 \lesssim T \lesssim T_6 $ (though we note that phonon hydrodynamics above a Debye temperature might look quite different to our model: in our model, $T_{\mathrm{BG}}$ is below the Debye temperature).

We propose studying plasmon dispersion relations in both of these materials, as the decreasing real part of the dispersion with temperature is a crisp signature of our electron-phonon hydrodynamics.  At higher temperatures, it may also be possible to carry out non-local magnetotransport experiments (similar to \cite{Berdyugin_2019} -- see the next section) to crisply test our theory.

\section{Magnetic fields}\label{sec:mag}
In this section, we describe the hydrodynamic coefficients upon turning on a relatively small, classical background magnetic field.  For simplicity, we restrict the discussion to two-dimensional fluids, applying a magnetic field perpendicular to the plane. The Boltzmann equation reads
\be
\pa_t |\Phi\ra+\b{F}_{\text{mag}}\cdot\nabla_{\b{p}} |\Phi\ra+W|\Phi\ra=E_i|J_i\ra,
\ee
where $\b{F}_{\text{mag}}$ is the Lorentz force acting merely on the electron modes:
\be
 (F_{\text{mag}})_i=-eB\ep_{ij}v_j(p).
\ee
We introduce the cyclotron frequency
\be
\omega_c=\frac{eBv_{\mathrm{F}}}{p_{\mathrm{F}}}.
\ee
We define the collision integral $W_{\text{mag}}|\Phi\ra=\b{F}_{\text{mag}}\cdot\nabla_{\b{p}} |\Phi\ra$, and note that $\la \Phi_1|W_{\text{mag}}|\Phi_2\ra=-\la \Phi_2|W_{\text{mag}}|\Phi_1\ra$. Such antisymmetry of $W_{\text{mag}}$ implies that these effects are, in some sense, dissipationless -- however, dissipative transport coefficients can and do become dependent on $B$.

\subsection{Viscosity}
Since the magnetic field necessarily breaks momentum conservation, we must actually consider a ``quasihydrodynamic" \cite{Grozdanov:2018fic} limit where the magnetic field is small: $\ga^{-1}\ll \omega_c^{-1}$ such that the momentum is still a long-lived and approximately conserved quantity.
The ``angle" dependence of $W_{\text{mag}}$ on a random Ansatz follows
\be
\la 0,m'|W_{\text{mag}}|0,m\ra_{\e}=\int^{2\pi}_0\frac{\ud \theta}{2\pi}e^{-\i m'\theta} (-eB\ep_{ij}v_j\frac{\pa}{\pa p_i})e^{\i m\theta} =\i m \omega_c \d_{m,m'}.
\ee
The shear viscosity \eqnref{eq:shear} is modified by the magnetic field through 
\be
\beal
\eta&\approx \frac{\nu\mu^2}{2}\left(\frac{1}{\ga_{\e}+2\i \omega_c}+\frac{1}{\ga_{\e}-2\i \omega_c}\right)+w^2v_{\ph}^2\left(\ga_{\ph}+\ga_{\text{ph-ph}}\right)^{-1}\\
&\approx\left\{ \begin{array}{ll}
\displaystyle \nu\mu^2 \left(\frac{\ga^{-1}}{1+(2\omega_c\ga^{-1})^2}+r^2\bw^2(\ga_{\ph}+\ga_{\mathrm{ph-ph}})^{-1}\right)  & T<T_{\text{BG}}\\
\displaystyle \nu\mu^2 \left(\frac{\bar{\gamma}^{-1}}{1+(2\omega_c\bar{\gamma}^{-1})^2}+r^2\bw^2(\ga_{\ph}+\ga_{\mathrm{ph-ph}})^{-1} \right)  & T_{\text{BG}}<T\ll T_{\mathrm{F}}\\
\end{array} \right. .
\eeal
\ee
However, if the phonon-phonon scattering is weak in the sense of $D_{\ph}\ll D_{\text{e-ph}}$, then the second term in the above equation could contribute non-trivially, making the electron-phonon fluid under background magnetic field an exotic fluid with unconventional shear viscosity.  This is analogous, in many ways, to the emergence of incoherent conductivities spoiling the conventional Kohn's theorem \cite{Hartnoll:2007ih}.  The Hall viscosity is given by
\be
\beal
\eta_{\text{H}}&=(\la\tau_{xx}|-\la\tau_{yy}|)\ga^{-1}_{ii}(|\tau_{xy}\ra+|\tau_{yx}\ra)\\
&\approx \frac{1}{2\i}\nu\mu^2\left(\frac{1}{\ga_{\e}+2\i \omega_c}-\frac{1}{\ga_{\e}-2\i \omega_c}\right)\\
&\approx\left\{ \begin{array}{ll}
\displaystyle -\nu\mu^2 \frac{2\omega_c\ga^{-2}}{1+(2\omega_c\ga^{-1})^2}  &T<T_{\text{BG}}\\
\displaystyle -\nu\mu^2 \frac{2\omega_c\bar{\gamma}^{-2}}{1+(2\omega_c\bar{\gamma}^{-1})^2}   & T_{\text{BG}}<T\ll T_{\mathrm{F}}\\
\end{array} \right. ,
\eeal
\ee
and it takes the more conventional form \cite{scaffidi} in a Fermi liquid.   Therefore, a simple test for the phonon contribution to transport and viscosity in an electron-phonon fluid would be to study the ratio \begin{equation}
    \mathcal{R}(B) = \frac{\eta^2 + \eta_{\mathrm{H}}^2}{\eta},
\end{equation}
which will be a decreasing function of $B$ because of the phonon contribution to shear viscosity.

Note that there are highly quantum effects which we have not captured in our semiclassical treatment \cite{luca}.   

\subsection{Magnetotransport}
\begin{figure}[t]
 \includegraphics[width=.8\linewidth]{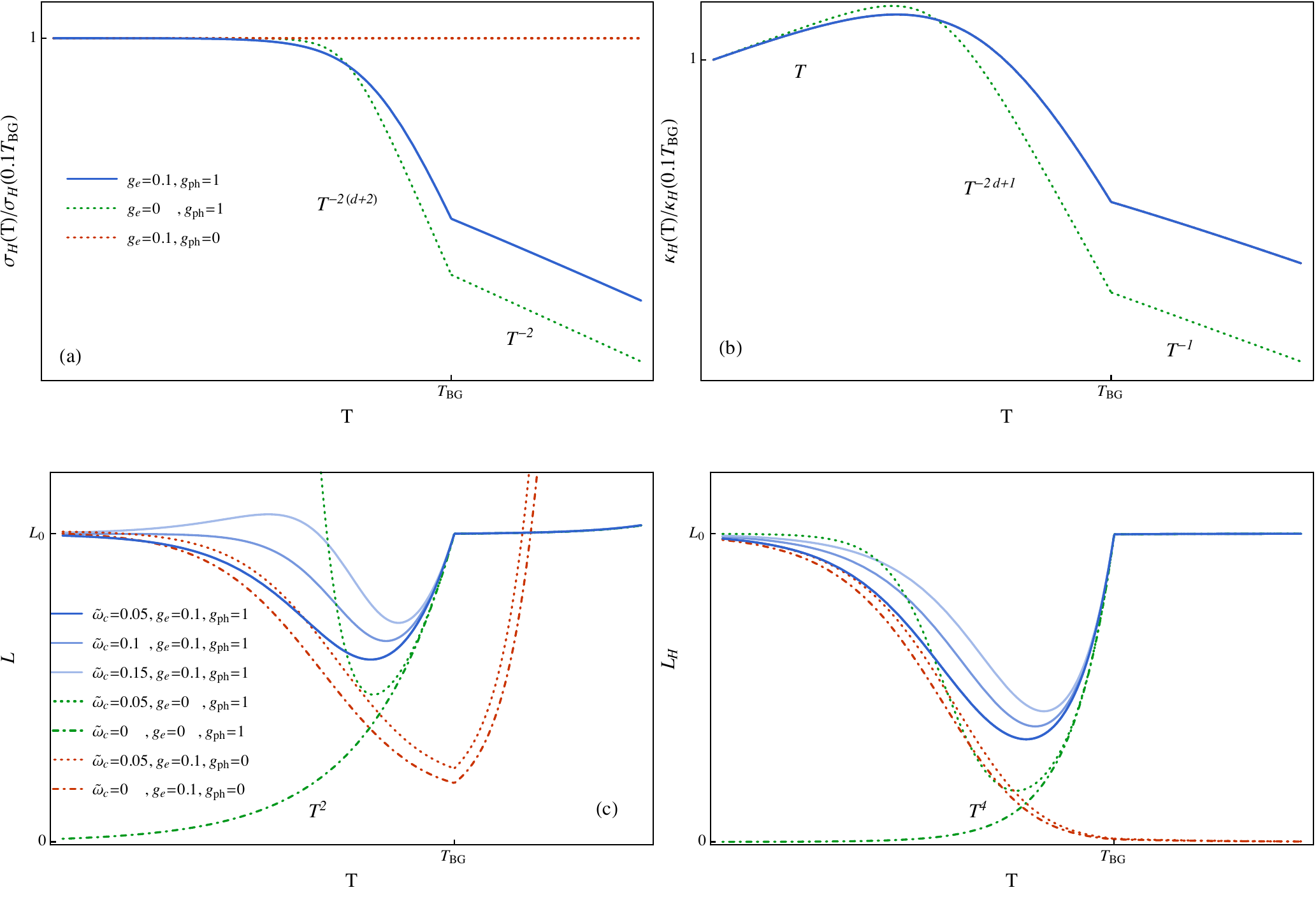}
\caption{Hall transport versus temperature with different field strengths and impurity scattering rates. (a-b) Log-log plot of electrical and thermal Hall conductivity against temperature with $\tilde{\omega}_c = 0.05$. $\kappa_{\mathrm{H}}$ does not depend on $g_\ph$. (c-d) Log-linear plot of (Hall) Lorenz number against temperature.
}
\label{fig:lorenzmag}
\end{figure}

We study the magnetotransport phenomenon by considering the strong electron phonon interaction, the impurities and the magnetic field together $\tilde{W}=W+W_{\text{imp}}+W_{\text{mag}}$. We present the results of conductivities in the limit $\Gamma_{\ph}\to 0$, and we will see that this limit preserves much of the interesting quantitative behavior of conductivities, but will simplify the expressions a lot. The complete expressions can be found in \appref{app:mag}. First, we find that the electrical conductivity is given by
\begin{subequations}
\begin{align}
    \sigma_{xx}&=e^2\frac{\nu}{2}v_{\mathrm{F}}^2\frac{\Gamma_{\e}}{\Gamma_{e}^2+\omega_c^2},\\
    \sigma_{yx}&=e^2\frac{\nu}{2}v_{\mathrm{F}}^2\frac{\omega_c}{\Gamma_{e}^2+\omega_c^2}.
\end{align}
\end{subequations}
They are very similar to the form in clean Fermi liquid \cite{PhysRevB.97.245128}. As discussed in the non-magnetic case, the $\sigma_{ij}$ become independent of electron-phonon interactions if $\Gamma_{\ph}$ is negligible, and coincide with the form in the non-interacting limit \eqnref{eqn:magnonint}. The open-circuit thermal conductivities are given by ($T<T_{\mathrm{BG}}$):
\begin{subequations}
\begin{align}
    T\kappa_{xx}&=\frac{\nu}{2}p_{\mathrm{F}}^2v_{\mathrm{F}}^4\left(\frac{\ba^2(\Gamma_{\e}+\gamma r^2/\ba^2)}{(\Gamma_{\e}+\gamma r^2/\ba^2)^2+\omega_c^2}+r^4\bw^4\gamma^{-1}\right),\\
    T\kappa_{yx}&=\frac{\nu}{2}p_{\mathrm{F}}^2v_{\mathrm{F}}^4\frac{\ba^2\omega_c}{(\Gamma_{\e}+\gamma r^2/\ba^2)^2+\omega_c^2},\\
\end{align}
\end{subequations}
while the closed-circuit thermal conductivities are 
\begin{subequations}
\begin{align}
    T\bar{\kappa}_{xx}&=T\kappa_{xx}+\frac{\nu}{2}p_{\mathrm{F}}^2v_{\mathrm{F}}^4\frac{r^4\bw^4\Gamma_{\e}}{\Gamma_{\e}^2+\omega_c^2},\\
    T\bar{\kappa}_{yx}&=T\kappa_{yx}+\frac{\nu}{2}p_{\mathrm{F}}^2v_{\mathrm{F}}^4\frac{r^4\bw^4\omega_c}{\Gamma_{\e}^2+\omega_c^2}.
\end{align}
\end{subequations}
We see that in the limit $B\to 0$, the closed-circuit thermal conductivity diverges in the clean limit $\Gamma_{\e}\to 0$ while the open-circuit thermal conductivity does not. As the Lorentz force does not act on phonon modes, one may naively think that phonon modes will not contribute to the Hall thermal conductivity. This is true for open-circuit Hall thermal conductivity, but not for closed-circuit:  in the latter case, the hybrid electron-phonon fluid is charged and participates in cyclotron motion, even when the energy density of the fluid is dominated by phonons.   The thermoelectric conductivity is given by ($T<T_{\mathrm{BG}}$)
\begin{subequations}
\begin{align}
    T\alpha_{xx}&=-e\frac{\nu}{2}p_{\mathrm{F}}v_{\mathrm{F}}^3\left(\frac{r^2\bw^2\Gamma_{\e}}{\Gamma_{\e}^2+\omega_c^2}+\frac{\ba^2(\Gamma_{\e}+\gamma r^2/\ba^2)}{(\Gamma_{\e}+\gamma r^2/\ba^2)^2+\omega_c^2}-\ba\frac{c\omega_c(2\Gamma_{\e}+\gamma r^2\ba^2)+(b-2\gamma r^2/\ba)(\Gamma_{\e}^2-\omega_c^2+\Gamma_{\e}\gamma r^2/\ba^2)}{((\Gamma_{\e}+\gamma r^2/\ba^2)^2+\omega_c^2)(\Gamma_{\e}+\omega_c^2)}\right),\\
    T\alpha_{yx}&=-e\frac{\nu}{2}p_{\mathrm{F}}v_{\mathrm{F}}^3\left( \frac{r^2\bw^2\omega_c}{\Gamma_{\e}+\omega_c^2}+\frac{\ba^2\omega_c}{(\Gamma_{\e}+\gamma r^2/\ba^2)^2+\omega_c^2}+\ba\frac{(2\gamma r^2/\ba-b)(2\Gamma_{\e}+\gamma r^2\ba^2)\omega_c+c(\Gamma_{\e}^2-\omega_c^2+\Gamma_{\e}\gamma r^2/\ba^2)}{((\Gamma_{\e}+\gamma r^2/\ba^2)^2+\omega_c^2)(\Gamma_{\e}+\omega_c^2)} \right).
\end{align}
\end{subequations}
where $c$ is the off-diagonal term in magnetic collision integral \eqnref{eqn:c}. For the Hall conductivities (in \appref{app:mag}), we show their temperature dependence in \figref{fig:lorenzmag}, where we defined the dimensionless parameter
\begin{equation}
    \tilde{\omega}_c=\frac{\omega_c}{r\gamma_0}.
\end{equation}
The behavior is quite similar to the dissipative conductivities but with the scaling sort of ``doubled'': the diagonal matrix element is squared in the determinant through the off-diagonal matrix element. Besides, the electric Hall conductivity won't diverge at clean limit $g_\e\to 0$ due to the cyclotron motion; the thermal Hall conductivity decreases as $T^{-1}$ above $T_{\mathrm{BG}}$.

We summarize the main results by studying two Lorenz numbers. The normal Lorenz number with magnetic field is given by (including $\Gamma_{\ph}$)  
\begin{equation}\label{eqn:maglorenz}
 L=\frac{\kappa_{xx}}{T\sigma_{xx}}\approx \left\{ \begin{array}{ll}
\displaystyle L_0 \left(  \frac{(\Ge+\ga r^2/\ba^2)}{(\Ge+\ga r^2/\ba^2)^2+\o^2}  + \frac{r^4\bw^2/\ba^2}{\Gamma_{\ph}+\gamma/\bw^2}  \right)\frac{(\Ge(\Gph+\ga/\bw^2)+\Gph\ga)^2+(\Gph+\ga/\bw^2)^2\o^2}{(\Gph+\ga/\bw^2)(\Ge(\Gph+\ga/\bw^2)+\Gph\ga)}  & ~ T<T_{\text{BG}}\\
\displaystyle L_0 \left(  \frac{(\Ge+\bar{\ga})}{(\Ge+\bar{\ga})^2+\o^2}  + \frac{r^4\bw^2/\ba^2}{\Gamma_{\ph}+\bar{\ga}/\bw^2}  \right)\frac{(\Ge(\Gph+\bar{\ga}/\bw^2)+\Gph\bar{\ga})^2+(\Gph+\bar{\ga}/\bw^2)^2\o^2}{(\Gph+\bar{\ga}/\bw^2)(\Ge(\Gph+\bar{\ga}/\bw^2)+\Gph\bar{\ga})} & ~T_{\text{BG}}<T\ll T_{\mathrm{F}}
 \end{array} \right. ,
\end{equation}
while the ``Hall'' Lorenz number is
\begin{equation}\label{eqn:maglorenz}
L_{\mathrm{H}}=\frac{\kappa_{yx}}{T\sigma_{yx}}\approx \left\{ \begin{array}{ll}
  \displaystyle L_0 \frac{(\Ge(\Gph+\ga/\bw^2)+\Gph\ga)^2+(\Gph+\ga/\bw^2)^2\o^2}{\left((\Ge+\ga r^2/\ba^2)^2+\omega_c^2\right)(\Gph+\ga/\bw^2)^2}& ~ T<T_{\text{BG}}\\
    \displaystyle L_0 \frac{(\Ge(\Gph+\bar{\ga}/\bw^2)+\Gph\bar{\ga})^2+(\Gph+\bar{\ga}/\bw^2)^2\o^2}{\left((\Ge+\bar{\ga})^2+\omega_c^2\right)(\Gph+\bar{\ga}/\bw^2)^2} & ~ T_{\text{BG}}<T\ll T_{\mathrm{F}}
\end{array} \right. .
\end{equation}
At low temperature, $L$ shows a non-monotonic temperature dependence when $\omega_{c}/\Gamma_\e>1$. This is similar to the result of interacting Fermi liquid \cite{PhysRevB.97.245128}, even though the underlying scattering mechanism is quite different. A nonzero $B$ field gives rise to a divergence for $g_\e=0$ while shows little influence for $g_\ph=0$, since in a hybrid electron-phonon fluid, only electrons directly interact with the $B$ field. 

As shown in Figure \ref{fig:lorenzmag}, the non-dissipative $L_{\mathrm{H}}$ strictly decreases with increasing temperature at low temperatures, in contrast to $L$.  It is important to note that, in principle, we can study $L_{\mathrm{H}}$ even as $B\rightarrow 0$: although $\sigma_{xy}$ and $\kappa_{xy}$ individually become very small, their ratio is fixed.  Interestingly, we find that if $g_\ph =0$, $L_{\mathrm{H}}$ continues to decrease at higher temperatures.  However, in most experimental systems, we expect that $g_\ph>0$, in which case we predict that near $T\sim T_{\mathrm{BG}}$, $L_{\mathrm{H}}\approx L_0$.

\section{Conclusion}\label{sec:discussion}
We have revisited the hydrodynamics and the thermoelectric transport of a fluid of coupled electrons and acoustic phonons in the presence of relatively strong electron-phonon interactions.  Solving the quantum Boltzmann equation, we found 9 different temperature regimes when $d>2$, and 7 temperature regimes for $d=2$, most of which exhibited unconventional thermodynamic properties and/or hydrodynamic transport coefficients such as viscosity or incoherent conductivity. 

An explicit and important example of unusual temperature dependence that we discovered occurs in the fluid's sound velocity.  In an intermediate temperature regime $T_4<T<T_6$, $v_{\mathrm{s}}\approx v_{\mathrm{ph}}/\bw \propto T^{-(d+1)/2}$ decreases with temperature, and is a clear experimental signature for the strongly coupled electron-phonon fluid. 
In particular, our model implies that in two dimensions, the exotic sound wave above will lead to an unusual plasmon dispersion relation, even in the presence of unscreened (or weakly screened) long-range Coulomb interactions.   The anomalous scaling of both the real and imaginary parts of $\omega_{\text{plasmon}}\sim (v_{\mathrm{ph}}/\bw) \sqrt{k}-\i\ga^{-1}|k| $,  represent unambiguous predictions that appear to us to be unique and ``smoking gun" signatures for electron-phonon hydrodynamics.  Since the real part of plasmon dispersion relations is quite readily measurable, we hope that our theory can be immediately tested with present day experimental capabilities (should an appropriate experimental candidate arise).

Another experimental test which might be possible (indirectly) with present day experimental techniques is to look for the unusual relationships between shear viscosity and Hall viscosity, as both of these coefficients can indirectly be measured using non-local transport experiments \cite{Levitov_2016,Torre_2015,pellegrino}.  Such experiments have been successfully performed in graphene \cite{Bandurin_2016,Berdyugin_2019}.  We hope that similar indirect measurements for incoherent conductivity can soon be developed.

After including electron-impurity and phonon-impurity scattering rates, we discussed the more conventionally studied thermoelectric transport coefficients of the bulk material.  As is well known \cite{PhysRevB.99.085104}, we obtain  a breakdown of the Wiedemann-Franz law at an intermediate temperature regime.  While this effect is \emph{not unique} to hydrodynamic theories of transport, it does represent a key signature for what temperature regime an experimental device may be operating in.  In many materials including $\mathrm{WTe}_2$, $\mathrm{WP}_2$ and $\mathrm{PtSn}_4$, notable dips in $L/L_0$ have appeared at quite low temperatures $\sim 30$ K.  Perhaps this corresponds in these materials to the temperature range $T\sim T_4$, where the conventional electron-phonon hydrodynamics of the literature -- which is entirely dominated by electrons, with phonons modifying only scattering rates -- is not applicable?  Especially in materials where first principles calculations \cite{coulter} are possible, it will be important to carefully estimate the temperature scales $T_1,\ldots,T_8$ in order to carefully match our theoretical predictions to future experiments.


A promising platform in realizing the simplest electron-phonon hydrodynamics developed in this paper should (at least) satisfy the conditions of low electron density and high mobility. The low density of electrons seems to demand a small Fermi surface, so that umklapp scattering is weak. At the same time, the impurity density must also be quite low, to facilitate a high mobility sample with very long momentum-relaxing mean free paths. They both contribute to the dominant momentum-conserving electron-phonon interaction. Moreover, the low density semimetal/semiconductor makes itself different from normal metal that $T_{\mathrm{BG}}$ drops from $\sim 300$ K to $\sim 30$ K. It is thus possible to have $T_{\mathrm{BG}}<T_{\mathrm{D}}$, where $T_{\mathrm{D}}$ is the Debye temperature, such that the phonon-drag peak temperature $\sim T_{\mathrm{D}}/10$ \cite{fu2018thermoelectric} is more likely to be located inside the highlighted regime $T_4<T<T_6$, indicating an out-of-equilibrium phonon mode. Such feature has been widely used in experiments \cite{fu2018thermoelectric,PhysRevB.85.035135} to suggest a strong electron-phonon interaction. Meanwhile, twisted bilayer graphene  \cite{PhysRevB.99.085105,PhysRevLett.90.056806,Polshyn_2019} is also a possible candidate system with strong electron-phonon interactions, although in these twisted samples, the ratio $r$ (which controlled much of the interesting physics) may become rather large.   Lastly, a system where very strong electron-electron and electron-phonon interactions might both coexist is in the Si MOSFETs \cite{kravchenko} where both $T_{\mathrm{F}}$ and $T_{\mathrm{BG}}$ can be below 30 K.  Since the hybrid electron-phonon sound speed does not depend on scattering rates, being intrinsically thermodynamic, such systems may also exhibit this unusual behavior.   As noted previously, the main limitation to studying electron-phonon hydrodynamics in these systems is the sample thickness; we hope that such a materials-dependent limitation may be overcome in future experiments.

Our work demonstrates an exciting possibility of uncovering novel hydrodynamic phenomena and plasmonics in quantum materials with correlated electrons and phonons.   We encourage experimentalists to carefully study plasmon dispersion relations in ultra-thin films of a number of compounds including $\mathrm{PtSn}_4$, $\mathrm{PdCoO}_2$, $\mathrm{WTe}_2$ and $\mathrm{WP}_2$.    In each of these materials, electron-phonon scattering has been argued to play a critical role in possibly unconventional transport physics.  However, all of these materials exhibit anisotropic Fermi surfaces and, as in \cite{caleb,varnavides2020generalized}, this anisotropy may well cause qualitative changes to our theory.  This is an important issue which we hope to address in the near future.


\section*{Acknowledgements}
We thank Sankar Das Sarma for useful feedback on the manuscript.  AL was supported by a Research Fellowship from the Alfred P. Sloan Foundation.

\appendix
\section{Normalization factor of rotationally invariant basis}\label{app:norm}
Following the definition in \eqnref{eqn:tildebasis}, we calculate the normalization for $n\leq 1$ to the leading order $T/T_{\mathrm{F}}$. The higher $n$ modes are suppressed by even higher powers of $T/T_{\mathrm{F}}$. They are
\bes
\begin{alignat}{5}
\la\tilde{0},\b{m}'|\tilde{0},\b{m}\ra_{\e}&=\int \frac{\ud^d p}{(2\pi\hbar)^d} \mathrm{Y}_{\b{m}}\mathrm{Y}_{\b{m}'}\left(-\frac{\pa f_{\mathrm{F}}}{\pa\ep}\right)|_p
=\Omega_{d-1}\d_{\b{m},\b{m}'}\int \frac{\ud \ep}{(2\pi\hbar)^dv_{\mathrm{F}}}p^{d-1} \d(\ep-\mu)\\
&=\nu\d_{\b{m},\b{m}'} +\mO\left(\frac{T^2}{T_{\mathrm{F}}^2}\right),\nonumber\\
\la\tilde{1},\b{m}'|\tilde{1},\b{m}\ra_{\e}&=\int \frac{\ud^d p}{(2\pi\hbar)^d}\mathrm{Y}_{\b{m}}\mathrm{Y}_{\b{m}'} (p-p_{\mathrm{F}})^2\left(-\frac{\pa f_{\mathrm{F}}}{\pa\ep}\right)|_p=\Omega_{d-1}\d_{\b{m},\b{m}'}\int \frac{\ud \ep}{(2\pi\hbar)^dv_{\mathrm{F}}}~p^{d-1} (p-p_{\mathrm{F}})^2\frac{\pi^2T^2}{3}\d''(\ep-\mu)\\
&=\nu\frac{\pi^2T^2}{3v_{\mathrm{F}}^2}\d_{\b{m},\b{m}'} +\mO\left(\frac{T^4}{T_{\mathrm{F}}^4}\right), \nonumber\\
\la\tilde{2},\b{m}'|\tilde{2},\b{m}\ra_{\e}&=\int \frac{\ud^d p}{(2\pi\hbar)^2}\mathrm{Y}_{\b{m}}\mathrm{Y}_{\b{m}'}  (p-p_{\mathrm{F}})^4\left(-\frac{\pa f_{\mathrm{F}}}{\pa\ep}\right)|_p
=\Omega_{d-1}\d_{\b{m},\b{m}'}\int \frac{\ud \ep}{(2\pi\hbar)^dv_{\mathrm{F}}}~p^{d-1} (p-p_{\mathrm{F}})^4\frac{7\pi^4T^4}{180}\d^{(4)}(\ep-\mu)\\
&=\mO\left(\frac{T^4}{T_{\mathrm{F}}^4}\right), \nonumber\\
\la\tilde{0},\b{m}'|\tilde{1},\b{m}\ra_{\e}&=\int \frac{\ud^d p}{(2\pi\hbar)^2}\mathrm{Y}_{\b{m}}\mathrm{Y}_{\b{m}'}  (p-p_{\mathrm{F}})\left(-\frac{\pa f_{\mathrm{F}}}{\pa\ep}\right)|_p
=\Omega_{d-1}\d_{\b{m},\b{m}'}\int \frac{\ud \ep}{(2\pi\hbar)^dv_{\mathrm{F}}}~p^{d-1} (p-p_{\mathrm{F}})\frac{\pi^2T^2}{3}\d''(\ep-\mu)\\
&=\frac{\nu}{2}\frac{\pi^2T^2}{3}(-\frac{\pa_pv_{\mathrm{F}}}{v_{\mathrm{F}}^2})\d_{\b{m},\b{m}'}+\mO\left(\frac{T^4}{T_{\mathrm{F}}^4}\right),\nonumber\\
\la\tilde{1},\b{m}'|\tilde{1},\b{m}\ra_{\ph}&=\int \frac{\ud^d q}{(2\pi\hbar)^d}\mathrm{Y}_{\b{m}}\mathrm{Y}_{\b{m}'}  q^2 \left(-\frac{\pa f_{\mathrm{B}}}{\pa\ep}\right)|_q
=\frac{\Omega_{d-1} T^{d+1}\d_{\b{m},\b{m}'}}{(2\pi\hbar)^dv_{\mathrm{ph}}^{d+2}}\int \ud x \frac{x^{d+1}e^{x}}{(e^x-1)^2}=2w^2\d_{\b{m},\b{m}'}, 
\end{alignat}
\ees
where we define
\begin{equation}\label{eqn:ph-norm}
   I(d) = \int\limits_0^\infty \ud x \frac{x^{d+1}e^{x}}{(e^x-1)^2}. 
\end{equation}


\section{Explicit expressions of electron-phonon relaxation rate}\label{app:e-ph}
First, we compute the collision integral after linearization. The zeroth-order of \eqnref{e-ph-coll} vanishes due to the detailed balance condition. Notice that 
\be
-\frac{\pa f^0_{\mathrm{F}}}{\pa\ep}=f^0_{\mathrm{F}}(1-f^0_{\mathrm{F}}),\quad -\frac{\pa f^0_{\mathrm{B}}}{\pa\ep}=b^0_{\mathrm{B}}(1+b^0_{\mathrm{B}}).
\ee
Then the first-order of $\mC_{\text{ph-e}}$ becomes
\be
\beal
T\d \mC_{\text{e-ph}}=&\left\{f_{\mathrm{F}k_2}(1-f_{\mathrm{F}k_2})(1-f_{\mathrm{F}k_1})(1+b_{\mathrm{B}q})+f_{\mathrm{F}k_1}f_{\mathrm{F}k_2}(1-f_{\mathrm{F}k_2})b_{\mathrm{B}q}\right\}\Phi_{k_2}\\
&+\left\{-f_{\mathrm{F}k_2}f_{\mathrm{F}k_1}(1-f_{\mathrm{F}k_1})(1+b_{\mathrm{B}q})-f_{\mathrm{F}k_1}(1-f_{\mathrm{F}k_1})(1-f_{\mathrm{F}k_2})b_{\mathrm{B}q}\right\}\Phi_{k_1}\\
&+\left\{f_{\mathrm{F}k_2}(1-f_{\mathrm{F}k_1})b_{\mathrm{B}q}(1+b_{\mathrm{B}q})-f_{\mathrm{F}k_1}(1-f_{\mathrm{F}k_2})b_{\mathrm{B}q}(1+b_{\mathrm{B}q})\right\}\Phi_{q}\\
=&(1-f_{\mathrm{F}k_2})f_{\mathrm{F}k_1}b_{\mathrm{B}q}(\Phi_{k_2}-\Phi_{k_1})+(f_{\mathrm{F}k_2}-f_{\mathrm{F}k_1})b_{\mathrm{B}q}(1+b_{\mathrm{B}q})\Phi_{q}\\
=&(1-f_{\mathrm{F}k_2})f_{\mathrm{F}k_1}b_{\mathrm{B}q}(\Phi_{k_2}-\Phi_{k_1}-\Phi_{q}),
\eeal
\ee
where in the last equation, we use the energy conservation.  \eqnref{S} is obtained straightforwardly. 

Next, we want to calculate the scaling on temperature of the collision integral. Based on the discussion below \eqnref{S}, we separate our discussions into two temperature regimes: $T<T_{\text{BG}}$ and $T>T_{\text{BG}}$.

\subsection{$T<T_{\text{BG}}$}
We evaluate \eqnref{S} in a general ground by assuming $\Phi$'s being $\ep$-dependent.
\be
\beal
\la\Phi| W_{\text{e-ph}} |\Phi\ra\approx&  \beta\int_{\b{q},\b{k}_1,\b{k}_2}|\L|^2\d(\b{k}_2-\b{k}_1-\b{q})\d(\ep_{k_2}-\ep_{k_1}-\omega_q)(1-f_{\mathrm{F}k_2})f_{\mathrm{F}k_1}f_{\mathrm{B}q}|\Phi|^2\\
=& \int_{\b{q}}  \int \frac{\ud \ep_2}{|v(\b{k}_2)|}\frac{\ud^{d-1} k_{\parallel}}{(2\pi)^d} \beta|\L|^2\d(\ep_{k_2}-\ep_{k_2-q}-\omega_q) (1-f_{\mathrm{F}k_2})f_{\mathrm{F}k_2-q}f_{\mathrm{B}q}|\Phi|^2\\
=&\int_{\b{q}}  \int \frac{\ud \ep_2}{|v(\b{k}_2)|}\frac{\ud^{d-1} k_{\parallel}}{(2\pi)^d} |\L|^2\d(\ep_{k_2}-\ep_{k_2-q}-\omega_q)\beta(1-f_{\mathrm{F}k_2})f_{\mathrm{F}k_2}\frac{f_{\mathrm{F}k_2-q}}{f_{\mathrm{F}k_2}}f_{\mathrm{B}q}|\Phi|^2\\
=&\int_{\b{q}}  \int \frac{\ud \ep_2}{|v(\b{k}_2)|}\frac{\ud^{d-1} k_{\parallel}}{(2\pi)^d} |\L|^2\d(\ep_{k_2}-\ep_{k_2-q}-\omega_q)(\d(\ep_2-\mu)+\frac{\pi^2T^2}{3}\d''(\ep_2-\mu)+...)\frac{f_{\mathrm{F}k_2-q}}{f_{\mathrm{F}k_2}}f_{\mathrm{B}q}|\Phi|^2\\
=&\int_{\b{q}}  \int \frac{\ud^{d-1} k_{\parallel}}{(2\pi)^dv_{\mathrm{F}}} D_{\text{e-ph}}|\b{q}|\frac{\d(\cos\th_{\b{q} \b{k}_{\parallel}}-v_{\mathrm{ph}}/v_{\mathrm{F}})}{v_{\mathrm{F}} q}\frac{2}{e^{-\beta\omega}+1}\frac{1}{e^{\beta\omega}-1}  \int \ud \ep_2(\d(\ep_2-\mu)+\frac{\pi^2T^2}{3}\d''(\ep_2-\mu)+...)|\Phi|^2\\
=&D_{\text{e-ph}}\int_{\b{q}}  \int \frac{\ud^{d-1} k_{\parallel}}{(2\pi)^dv_{\mathrm{F}}^2} \d(\cos\th_{\b{q} \b{k}_{\parallel}}-v_{\mathrm{ph}}/v_{\mathrm{F}})\frac{1}{\sinh(\beta\omega)} \int \ud \ep_2(\d(\ep_2-\mu)+\frac{\pi^2T^2}{3}\d''(\ep_2-\mu)+...)|\Phi|^2\\
\approx&D_{\text{e-ph}}\int_{\b{q}}  \int \frac{\ud^{d-1} \b{k}_{\parallel}}{(2\pi)^dv_{\mathrm{F}}^2} \d(\cos\th_{\b{q} \b{k}_{\parallel}}-v_{\mathrm{ph}}/v_{\mathrm{F}})\times \Theta(T-v_{\mathrm{ph}} q)\int \ud \ep_2(\d(\ep_2-\mu)+\frac{\pi^2T^2}{3}\d''(\ep_2-\mu)+...)|\Phi|^2\\
=&D_{\text{e-ph}}\int_0^{T/v_{\mathrm{ph}}}\frac{\ud^d q}{(2\pi)^d}\int \frac{\ud^{d-1} k_{\parallel}}{(2\pi)^dv_{\mathrm{F}}^2}\d(\cos\th_{\b{q}\b{k}_{\parallel}}-v_{\mathrm{ph}}/v_{\mathrm{F}})\int \ud \ep_2(\d(\ep_2-\mu)+\frac{\pi^2T^2}{3}\d''(\ep_2-\mu)+...)|\Phi|^2\\
\eeal
\label{intS}
\ee
We find that $\th_{\b{q}\b{k}_{\parallel}}\approx \pi/2$ implying that phonon momentum is approximately perpendicular to electron momentum. 
We then apply the ``Bloch Ansatz" $|\Phi\ra_{\ph}=0$, and replacing the $|\Phi\ra$ mode with rotationally invariant basis. We have
\bes
\be
\beal
_{\e}\la \tilde{0}|W_{\text{e-ph}}|\tilde{0}\ra_{\e}
\approx D_{\text{e-ph}}\int_0^{T/v_{\mathrm{ph}}}\frac{\ud^d q}{(2\pi)^d}\int \frac{\ud^{d-1} k_{\parallel}}{(2\pi)^dv_{\mathrm{F}}^2}\d(\cos\th_{\b{q}\b{k}_{\parallel}}-v_{\mathrm{ph}}/v_{\mathrm{F}})\frac{q^2}{p_{\mathrm{F}}^2}
= \a^2(2) \frac{1}{p_{\mathrm{F}}^2} T^{d+2},
\eeal
\ee
\be
\beal
_{\e}\la \tilde{1}|W_{\text{e-ph}}|\tilde{1}\ra_{\e}
\approx  D_{\text{e-ph}}\int_0^{T/v_{\mathrm{ph}}}\frac{\ud^d q}{(2\pi)^d}\int \frac{\ud^{d-1} k_{\parallel}}{(2\pi)^dv_{\mathrm{F}}^2}\d(\cos\th_{\b{q}\b{k}_{\parallel}}-v_{\mathrm{ph}}/v_{\mathrm{F}})\frac{v_{\mathrm{ph}}^2}{v_{\mathrm{F}}^2}q^2
=  \a^2(2) \frac{v_{\mathrm{ph}}^2}{v_{\mathrm{F}}^2} T^{d+2},
\eeal
\ee
\be
\beal
_{\e}\la \tilde{2}|W_{\text{e-ph}}|\tilde{2}\ra_{\e}
\approx  D_{\text{e-ph}}\int_0^{T/v_{\mathrm{ph}}}\frac{\ud^d q}{(2\pi)^d}\int \frac{\ud^{d-1} k_{\parallel}}{(2\pi)^dv_{\mathrm{F}}^2}\d(\cos\th_{\b{q}\b{k}_{\parallel}}-v_{\mathrm{ph}}/v_{\mathrm{F}})\frac{v_{\mathrm{ph}}^4}{v_{\mathrm{F}}^4}q^4
=   \a^2(4) \frac{v_{\mathrm{ph}}^4}{v_{\mathrm{F}}^4} T^{d+4},
\eeal
\ee
\ees
where 
\begin{equation}
    \a^2(n)=D_{\text{e-ph}}\int_0^1\frac{\ud^d x x^{n}}{(2\pi)^d}\int \frac{\ud^{d-1} k_{\parallel}}{(2\pi)^dv_{\mathrm{F}}^2v_{\ph}^{d+n}}\d(\cos\th_{\b{x}\b{k}_{\parallel}}-v_{\ph}/v_{\mathrm{F}}).
\end{equation}
To derive the relaxation time for phonons, we set $\Phi_{\e}=0$. Then the collision integral for phonon modes is given by
\be
\beal
_{\ph}\la \tilde{1}|W_{\text{e-ph}}|\tilde{1}\ra_{\ph}
\approx D_{\text{e-ph}}\int_0^{T/v_{\mathrm{ph}}}\frac{\ud^d q}{(2\pi)^d}\int \frac{\ud^{d-1} k_{\parallel}}{(2\pi)^dv_{\mathrm{F}}^2}\d(\cos\th_{\b{q}\b{k}_{\parallel}}-v_{\mathrm{ph}}/v_{\mathrm{F}})
(q\hat{q})^2
=\a^2(2)T^{d+2} .
\eeal
\ee
\subsection{$T>T_{\text{BG}}$}
We need to recalculate the collision integral \eqnref{intS}, and this time the phonon behaves more like a classical boson gas with equipartition distribution:
\be
\beal
\la\Phi| W_{\text{e-ph}}|\Phi\ra\approx &\int_{\b{q}}  \int \frac{\ud^{d-1} k_{\parallel}}{(2\pi)^dv_{\mathrm{F}}} |\L|^2\d(\ep_{k_2}-\ep_{k_2-q}-\omega_q)\frac{1}{\sinh(\beta\omega_q)} \int \ud \ep_2(\d(\ep_2-\mu)+\frac{\pi^2T^2}{3}\d''(\ep_2-\mu)+...)|\Phi|^2\\
\approx & D_{\text{e-ph}}\int^{k_F}\frac{\ud^d q}{(2\pi)^d}   \int \frac{\ud^{d-1} k_{\parallel}}{(2\pi)^dv_{\mathrm{F}}}  \d(\ep_{k_2}-\ep_{k_2-q}-\omega_q)\frac{T}{v_{\ph}} \int \ud \ep_2(\d(\ep_2-\mu)+\frac{\pi^2T^2}{3}\d''(\ep_2-\mu)+...)|\Phi|^2.\\
\eeal
\ee
 We apply $|\Phi\ra_{\ph}=0$, and find
\bes
\be
\beal
_{\e}\la \tilde{0}|W_{\text{e-ph}}|\tilde{0}\ra_{\e}
\approx D_{\text{e-ph}}\int^{k_F}\frac{\ud^d q}{(2\pi)^d}   \int \frac{\ud^{d-1} k_{\parallel}}{(2\pi)^dv_{\mathrm{F}}}  \d(\ep_{k_2}-\ep_{k_2-q}-\omega_q)\frac{T}{v_{\ph}} \frac{q^2}{p_{\mathrm{F}}^2}
=\a'^2(2) \frac{1}{p_{\mathrm{F}}^2}T,
\eeal
\ee
\be
\beal
_{\e}\la \tilde{1}|W_{\text{e-ph}}|\tilde{1}\ra_{\e}
\approx D_{\text{e-ph}} \int^{k_F}\frac{\ud^d q}{(2\pi)^d}   \int \frac{\ud^{d-1} k_{\parallel}}{(2\pi)^dv_{\mathrm{F}}}  \d(\ep_{k_2}-\ep_{k_2-q}-\omega_q)\frac{T}{v_{\ph}} a^2
\approx  \a'^2(2) \frac{a^2}{p_{\mathrm{F}}^2}T,
\eeal
\ee
\ees
where
\begin{equation}
    \a'^2(n)=D_{\text{e-ph}} \int^{k_{\mathrm{F}}}\frac{\ud^d q}{(2\pi)^d}   \int \frac{\ud^{d-1} k_{\parallel}}{(2\pi)^dv_{\mathrm{F}}}  \d(\ep_{k_{2}}-\ep_{k_{2}-q}-\omega_q)\frac{q^n}{v_{\ph}}.
\end{equation}
If we apply $|\Phi\ra_{\e}=0$, we have
\be
\beal
_{\ph}\la \tilde{1}|W_{\text{e-ph}}|\tilde{1}\ra_{\ph}
\approx D_{\text{e-ph}} \int^{k_F}\frac{\ud^d q}{(2\pi)^d}   \int \frac{\ud^{d-1} k_{\parallel}}{(2\pi)^dv_{\mathrm{F}}}  \d(\ep_{k_2}-\ep_{k_2-q}-\omega_q)\frac{T}{v_{\ph}} q^2
=\a'^2(2) T.
\eeal
\ee

\section{Matrix elements of collision integrals without magnetic field}\label{app:wf}


According to \eqnref{eqn:Weph}, we have
\be
W_{\text{e-ph}}' = \left( \begin{array}{ccc} 
\ba^2+\bar{w}^2 & -\bar{a} & -\bar{w} \\ 
-\bar{a} & 1+\bar{w}^2 & -\bar{a}\bar{w}  \\ 
-\bar{w} &-\bar{a}\bar{w} & 1+\ba^2 
\end{array} \right) 
\frac{W_{\text{e-ph}}}{(1+\ba^2+\bar{w}^2)^2}
\left( \begin{array}{ccc} 
\ba^2+\bar{w}^2 & -\bar{a} & -\bar{w} \\ 
-\bar{a} & 1+\bar{w}^2 & -\bar{a}\bar{w}  \\ 
-\bar{w} &-\bar{a}\bar{w} & 1+\ba^2 
\end{array} \right) .
\ee
We write 
\be
W=W_{\text{imp}}^\e+W_{\text{imp}}^{\ph}+W_{\text{e-ph}}'= \left( \begin{array}{ccc} 
A_1 & A_4 & A_5 \\ 
A_4 & A_2 & A_6 \\ 
A_5&A_6& A_3
\end{array} \right) ,
\ee
where for $T<T_{\text{BG}}$,
\begin{subequations}\begin{align}
A_1&=\Ge+\ga\frac{(1+r^2+(\ba^2+\bw^2)^2)}{(1+\ba^2+\bw^2)^2},\\
A_2&=\Ge+\ga\frac{r^2(1+\bw^2)^2}{\ba^2(1+\ba^2+\bw^2)^2},\\
A_3&=\Gph+\ga\frac{((1+\ba^2)^2+\bw^4)}{\bw^2(1+\ba^2+\bw^2)^2},\\
A_4&=b-\ga\frac{\ba^2(-1+\bw^2)+r^2(1+\bw^2)}{\ba(1+\ba^2+\bw^2)^2},\\
A_5&=-\ga\frac{1+(\ba^2-r^2)\bw^2+\bw^4}{\bw(1+\ba^2+\bw^2)^2},\\
A_6&=\ga\frac{\ba^2(-1+\bw^2)-r^2(1+\bw^2)}{\ba\bw(1+\ba^2+\bw^2)^2},
\end{align}\end{subequations}
and for $T>T_{\text{BG}}$,
\begin{subequations}\begin{align}
A_1&=\Ge+\bar{\gamma}\frac{(1+\ba^4+\bw^4+\ba^2(1+2\bw^2))}{(1+\ba^2+\bw^2)^2},\\
A_2&=\Ge+\bar{\gamma}\frac{2\ba^2+(1+\bw^2)^2}{(1+\ba^2+\bw^2)^2},\\
A_3&=\Gph+\bar{\gamma}\frac{(1+\ba^2)(1+\ba^2+\bw^4)}{\bw^2(1+\ba^2+\bw^2)^2},\\
A_4&=b-\bar{\gamma}\frac{\ba(\ba^2+2\bw^2)}{(1+\ba^2+\bw^2)^2},\\
A_5&=-\bar{\gamma}\frac{1+\ba^2+\bw^4}{\bw(1+\ba^2+\bw^2)^2},\\
A_6&=\bar{\gamma}\frac{\ba(1+\ba^2+\bw^4)}{\bw(1+\ba^2+\bw^2)^2}.
\end{align}\end{subequations}
The inverse of $W$ matrix is given by
\be
W^{-1} = D^{-1}
\left( \begin{array}{ccc} 
 A_2A_3-A_6^2&-A_3A_4+A_5A_6 &-A_2A_5+A_4A_6 \\
 -A_3A_4+A_5A_6 &A_1A_3-A_5^2 &A_4A_5-A_1A_6 \\
  -A_2A_5+A_4A_6 & A_4A_5-A_1A_6 &A_1A_2-A_4^2 \\
\end{array} \right) ,
\ee
where
\be
D=\det W=A_1A_2A_3-A_3A_4^2-A_2A_5^2+2A_4A_5A_6-A_1A_6^2.
\ee
We then approximate the $W^{-1}$ in the above two temperature regimes. 

At $T<T_{\text{BG}}$ we have the determinant
\begin{equation}
    D\approx (\Ge+r^2\ga/\ba^2)[\Ge(\Gph+\ga/\bw^2)+\Gph\ga],
\end{equation}
and the collision integral becomes
\be
W^{-1} \approx
 \left( \begin{array}{ccc} 
 \frac{\Gph+\ga/\bw^2}{\Ge(\Gph+\ga/\bw^2)+\Gph\ga}&  \frac{-(\Gph+\ga/\bw^2)(b-\ga r^2/\ba)+\ga^2r^2/(\ba\bw^2)}{D} & \frac{\ga/\bw}{\Ge(\Gph+\ga/\bw^2)+\Gph\ga} \\
 \frac{-(\Gph+\ga/\bw^2)(b-\ga r^2/\ba)+\ga^2r^2/(\ba\bw^2)}{D} & \frac{1}{\Ge+r^2\ga/\ba^2}& \frac{-(b-\ga r^2/\ba)\ga/\bw+(\Ge+\ga)\ga r^2/(\ba\bw)}{D}   \\
 \frac{\ga/\bw}{\Ge(\Gph+\ga/\bw^2)+\Gph\ga} &   \frac{-(b-\ga r^2/\ba)\ga/\bw+(\Ge+\ga)\ga r^2/(\ba\bw)}{D} & \frac{\Ge+\ga}{\Ge(\Gph+\ga/\bw^2)+\Gph\ga} \\
\end{array} \right) ,
\label{winv1}
\ee
while at $T>T_{\text{BG}}$ we have the determinant
\begin{equation}
     D\approx (\Ge+\bar{\gamma})[\Ge(\Gph+\bar{\gamma}/\bw^2)+\Gph\bar{\gamma}],
\end{equation}
and the collision integral becomes
\be
W^{-1} \approx
\left( \begin{array}{ccc} 
 \frac{\Gph+\bar{\gamma}/\bw^2}{\Ge(\Gph+\bar{\gamma}/\bw^2)+\Gph\bar{\gamma}}&\frac{-(\Gph+\bar{\gamma}/\bw^2)(b-\bar{\gamma}\ba)+\bar{\gamma}^2\ba/\bw^2}{D} &\frac{\bar{\gamma}/\bw}{\Ge(\Gph+\bar{\gamma}/\bw^2)+\Gph\bar{\gamma}} \\
\frac{-(\Gph+\bar{\gamma}/\bw^2)(b-\bar{\gamma}\ba)+\bar{\gamma}^2\ba/\bw^2}{D} &\frac{1}{\Ge+\bar{\gamma}}& \frac{-(b-\bar{\gamma}\ba)\bar{\gamma}/\bw+(\Ge+\bar{\gamma})\bar{\gamma}\ba/\bw}{D}   \\
\frac{\bar{\gamma}/\bw}{\Ge(\Gph+\bar{\gamma}/\bw^2)+\Gph\bar{\gamma}} &  \frac{-(b-\bar{\gamma}\ba)\bar{\gamma}/\bw+(\Ge+\bar{\gamma})\bar{\gamma}\ba/\bw}{D} & \frac{\Ge+\bar{\gamma}}{\Ge(\Gph+\bar{\gamma}/\bw^2)+\Gph\bar{\gamma}} \\
\end{array} \right) .
\label{winv2}
\ee
In above equations, we keep the leading order in $T/T_{\mathrm{F}}$.

\section{Transport with low phonon scattering}
\label{app:newregime}
When momentum relaxing phonon scattering is low ($g_{\mathrm{ph}}\to 0$), drastic changes show up in the temperature scaling of thermoelectric transport quantities (see \figref{fig:lorenz2}).  More precisely, consider the limit
\begin{equation}
    0<\Gamma_{\mathrm{ph}}\ll \gamma/\bw^2~ \Longrightarrow~ 0<g_{\mathrm{ph}}\ll T/T_{\mathrm{F}}.
\end{equation}
Hence, we obtain, when $T<T_{\mathrm{BG}}$,
\begin{subequations}
\begin{align}
    \sigma_{xx}(g_{\mathrm{ph}}\to 0)&\approx e^2\frac{\nu}{2}v_{\mathrm{F}}^2\frac{1}{\Gamma_{\e}+\Gamma_{\ph}\bw^2},\\
    T\kappa_{xx}(g_{\mathrm{ph}}\to 0)&\approx \frac{\nu}{2}p_{\mathrm{F}}^2v_{\mathrm{F}}^4\left(\frac{\ba^2}{\Gamma_{\e}+\ga r^2/\ba^2}+\frac{r^4\bw^4}{\ga} \right),\\
    L(g_{\mathrm{ph}}\to 0)/L_0&\approx \frac{\Ge+\Gph\bw^2}{\Ge+\ga r^2/\ba^2}+\frac{r^4\bw^4}{\ba^2}\frac{\Ge+\Gph\bw^2}{\ga}.
\end{align}
\end{subequations}
A similar calculation can be performed for $T>T_{\mathrm{BG}}$.

The electrical conductivity does not depend on the electron-phonon scattering strength $\ga$ now; it saturates to a constant when $g_{\mathrm{ph}}=0$. The saturation is the canonical Drude result in non-interacting electron systems \cite{PhysRevB.97.245128} (see also \figref{fig:lorenz2}(a)). The thermal conductivity is also affected by such ultra-clean lattice as depicted in \figref{fig:lorenz2}(b), where no plateau appears above $T_{\mathrm{BG}}$ but a faster scaling $\sim T^{2d+1}$; the scaling $\kappa\sim T^{-d+1}$ is also get corrected by the phonon contribution. Although $T^*$, which is the temperature where $L$ starts to deviate from $L_0$ as defined in \eqnref{eq:Tstar}, is roughly not affected, $T_{\mathrm{min}}=x T_{\mathrm{BG}}$ has changed and now the $x$ is given by the solution of
\begin{equation}
    -d r g_\e + (d+1)g_\ph g_\e x+ g_\ph x^{d+1}=0.
\end{equation}
Significantly, lowering $g_\ph$ does affect $L$ below $T_{\mathrm{BG}}$ --- $T_{\mathrm{min}}$ \textit{increases} --- contrasting the case in the main text (see \figref{fig:lorenz2}(c)).

\section{Matrix elements of collision integrals with magnetic field}\label{app:mag}
We consider only $|0_{x,y}\ra_{\e},|1_{x,y}\ra_{\e}$ and $|1_{x,y}\ra_{\ph}$ to keep track of the leading order temperature dependence of all conductivities.  Here the integers correspond to the radial modes (not the angular modes).  We focus on $d=2$ spatial dimensions. We write 
\be
\tilde{W}=W\otimes\d_{ij}+\tilde{W}_{\text{mag}}\otimes \ep_{ij},
\label{eq:magW}
\ee
where
\be
\tilde{W}_{\text{mag}}=\omega_c\left(|0\ra_\e\la 0|_\e+|1\ra_\e\la 1|_\e\right)+c\left(|0\ra_\e\la 1|_\e+|1\ra_\e\la 0|_\e\right) ,
\ee
and
\begin{equation}\label{eqn:c}
    c=\frac{\pi T eB(p_{\mathrm{F}}\pa_pv_{\mathrm{F}}-v_{\mathrm{F}})}{\sqrt{3}v_{\mathrm{F}}p_{\mathrm{F}}}.
\end{equation}
Using \eqnref{eq:magW}, we are allowed to invert the $6\times 6$ collision matrix to calculate the conductivities.

let's begin by studying the non-interacting limit $\gamma\to 0$. After some algebra, we have
\begin{subequations}\label{eqn:magnonint}
\begin{align}
\sigma_{xx}&\approx e^2\frac{\nu}{2}v_{\mathrm{F}}^2\frac{\Gamma_{\e}}{\Gamma_{\e}^2+\omega_{c}^2},\\
\sigma_{yx}&\approx e^2\frac{\nu}{2}v_{\mathrm{F}}^2\frac{\omega_{c}}{\Gamma_{\e}^2+\omega_{c}^2},\\
T\kappa_{xx}\approx T\bar{\kappa}_{xx}&\approx \frac{\nu}{2}p_{\mathrm{F}}^2v_{\mathrm{F}}^4 \left(\frac{\ba^2 \Gamma_{\e}}{\Gamma_{\e}^2+\omega_{c}^2}+\frac{r^4\bw^2}{\Gamma_{\ph}}  \right),\\
T\kappa_{yx}\approx T\bar{\kappa}_{yx}&\approx \frac{\nu}{2}p_{\mathrm{F}}^2v_{\mathrm{F}}^4\frac{\ba^2\omega_{c}}{\Gamma_{\e}^2+\omega_{c}^2},\\
T\a_{xx}&\approx -e\frac{\nu}{2}p_{\mathrm{F}}v_{\mathrm{F}}^3 \left( \frac{\ba^2 \Gamma_{\e}}{\Gamma_{\e}^2+\omega_{c}^2}-\ba\frac{b(\Gamma_{\e}^2-\omega_{c}^2)+2c\Gamma_{\e}\omega_{c}}{(\Gamma_{\e}^2+\omega_{c}^2)^2}  \right),\\
T\a_{yx}&\approx -e\frac{\nu}{2}p_{\mathrm{F}}v_{\mathrm{F}}^3 \left( \frac{\ba^2\omega_{c}}{\Gamma_{\e}^2+\omega_{c}^2}+\ba\frac{c(\Gamma_{\e}^2-\omega_{c}^2)-2b\Gamma_{\e}\omega_{c}}{(\Gamma_{\e}^2+\omega_{c}^2)^2}  \right).
\end{align}
\end{subequations}
In above equations, we keep the leading order in the limit $T\to 0$. They are well reduced to the non-magnetic cases as discussed in the main text. We see that only (non-Hall) thermal conductivity receives a phonon's contribution which is totally $B$-irrelevant. This leads to two consequences. First, for a large enough $B$, phonon modes will dominate the thermal conductivity; Second, in the clean limit $\Gamma_{\e,\ph}\to 0$, the divergence shows up
in thermal conductivity through the phonon modes. 

Including the momentum-conserving electron-phonon interaction, we obtain at $T<T_{\mathrm{BG}}$,
\begin{subequations}
\begin{align}
\sigma_{xx}&\approx  e^2\frac{\nu}{2}v_{\mathrm{F}}^2 \frac{(\ga+\Gph\bw^2)(\ga\Ge+(\ga+\Ge)\Gph\bw^2)}{\ga^2(\Ge^2+\o^2)+2\ga\Gph(\Ge(\ga+\Ge)+\o^2)\bw^2+\Gph^2((\ga+\Ge)^2+\o^2)\bw^4},\\
\sigma_{yx}&\approx e^2\frac{\nu}{2}v_{\mathrm{F}}^2 \frac{\o(\ga+\Gph\bw^2)^2}{\ga^2(\Ge^2+\o^2)+2\ga\Gph(\Ge(\ga+\Ge)+\o^2)\bw^2+\Gph^2((\ga+\Ge)^2+\o^2)\bw^4},\\
T\bar{\kappa}_{xx}&\approx \frac{\nu}{2}p_{\mathrm{F}}^2v_{\mathrm{F}}^4 \left(  \frac{(\ba^2\Ge+r^2\ga)}{\Ge^2+\o^2+2\Ge\ga r^2/\ba^2+\ga^2r^4/\ba^4}+ \frac{r^4\bw^2( \bw^2\ga[\Ge(\ga+\Ge)+\o^2]+\bw^4\Gph[(\ga+\Ge)^2+\o^2] )}{\ga^2(\Ge^2+\o^2)+2\ga\Gph(\Ge(\ga+\Ge)+\o^2)\bw^2+\Gph^2((\ga+\Ge)^2+\o^2)\bw^4} \right),\\
T\bar{\kappa}_{yx}&\approx  \frac{\nu}{2}p_{\mathrm{F}}^2v_{\mathrm{F}}^4 \left( \frac{\o\ba^2}{(\Ge^2+\o^2)+2\Ge\ga r^2/\ba^2+\ga^2r^4/\ba^4}  + \frac{r^4\bw^4\ga^2\o}{\ga^2(\Ge^2+\o^2)+2\ga\Gph(\Ge(\ga+\Ge)+\o^2)\bw^2+\Gph^2((\ga+\Ge)^2+\o^2)\bw^4} \right),\\
T\a_{xx}&\approx -e\frac{\nu}{2}p_{\mathrm{F}}v_{\mathrm{F}}^3 \Bigg(\frac{\ga r^2\bw^2(\ga\Ge+(\ga+\Ge)\Gph\bw^2)}{\ga^2(\Ge^2+\o^2)+2\ga\Gph(\Ge(\ga+\Ge)+\o^2)\bw^2+\Gph^2((\ga+\Ge)^2+\o^2)\bw^4}+ \frac{(\ba^2\Ge+r^2\ga)}{\Ge^2+\o^2+2\Ge\ga r^2/\ba^2+\ga^2r^4/\ba^4}\\
&  +\frac{\ba}{\left((\Gamma_{\e}+\ga r^2/\ba^2)^2+\omega_c^2\right)\left((\gamma\Gamma_{\e}+\Gamma_{\ph}(\gamma+\Gamma_{\e})\bw^2)^2+\omega_c^2(\gamma+\Gamma_{\ph}\bw^2)^2 \right)}\nonumber\\
&\times \Bigg\{(\ga+\Gamma_{\ph}\bw^2)\left[-c(2\gamma\Gamma_{\e}\omega_c+\Gamma_{\ph}(\gamma+2\Gamma_{\e})\omega_c\bw^2)+b(\gamma(-\Gamma_{\e}^2+\omega_c^2)+\Gamma_{\ph}(-\Gamma_{\e}(\gamma+\Gamma_{\e})+\omega_c^2)\bw^2)\right] \nonumber\\
& -\gamma r^2/\ba^2\left[ c\omega_c(\gamma+\Gamma_{\ph})^2+b(\gamma^2\Gamma_{\e}+\gamma\Gamma_{\ph}(\gamma+2\Gamma_{\e})\bw^2+\Gamma_{\ph}^2(\gamma+\Gamma_{\e})\bw^4 )    \right] \nonumber\\
& -\ba\gamma r^2/\ba^2 (2\gamma+\Gamma_{\ph}\bw^2)\left[\gamma(-\Gamma_{\e}^2+\omega_c^2)+\Gamma_{\ph}(-\Gamma_{\e}(\gamma+\Gamma_{\e})+\omega_c^2)\bw^2  \right] \nonumber\\
&+\ba\gamma^2r^4/\ba^4 (2\gamma+\Gamma_{\ph}\bw^2)\left[ \Gamma_{\e}\Gamma_{\ph}\bw^2+\gamma(\Gamma_{\e}+\Gamma_{\ph}\bw^2)\right]\Bigg\}\Bigg)\nonumber ,\\
T\a_{yx}&\approx -e\frac{\nu}{2}p_{\mathrm{F}}v_{\mathrm{F}}^3 \Bigg( \frac{\o\ga r^2\bw^2(\ga+\Gph\bw^2)}{\ga^2(\Ge^2+\o^2)+2\ga\Gph(\Ge(\ga+\Ge)+\o^2)\bw^2+\Gph^2((\ga+\Ge)^2+\o^2)\bw^4}+ \frac{\o\ba^2}{(\Ge^2+\o^2)+2\Ge\ga r^2/\ba^2+\ga^2r^4/\ba^4}\\
&+\frac{\ba}{\left((\Gamma_{\e}+\ga r^2/\ba^2)^2+\omega_c^2\right)\left((\gamma\Gamma_{\e}+\Gamma_{\ph}(\gamma+\Gamma_{\e})\bw^2)^2+\omega_c^2(\gamma+\Gamma_{\ph}\bw^2)^2 \right)}\nonumber \\
&\times \Bigg\{(\gamma+\Gamma_{\ph}\bw^2)\left[-b(2\gamma\Gamma_{\e}\omega_c+\Gamma_{\ph}(\gamma+2\Gamma_{\e})\omega_c\bw^2)-c(\gamma(-\Gamma_{\e}^2+\omega_c^2)+\Gamma_{\ph}(-\Gamma_{\e}(\gamma+\Gamma_{\e})+\omega_c^2)\bw^2) \right] \nonumber \\
&- \gamma r^2/\ba^2\left[ b\omega_c(\gamma+\Gamma_{\ph})^2-c(\gamma^2\Gamma_{\e}+\gamma\Gamma_{\ph}(\gamma+2\Gamma_{\e})\bw^2+\Gamma_{\ph}^2(\gamma+\Gamma_{\e})\bw^4 )    \right] \nonumber\\
&+ \ba\gamma r^2/\ba^2\omega_c (2\gamma+\Gamma_{\e}\bw^2)(2\gamma\Gamma_{\e}+(\gamma+2\Gamma_{\e})\Gamma_{\ph}\bw^2 )\nonumber\\
&+\ba\gamma^2 r^4/\ba^4 \omega_c (2\gamma+\Gamma_{\ph}\bw^2)(\gamma+\Gamma_{\ph}\bw^2)    \Bigg\}    \Bigg), \nonumber
\end{align}
\end{subequations}
and at  $T>T_{\mathrm{BG}}$,
\begin{subequations}
\begin{align}
\sigma_{xx}&\approx  e^2\frac{\nu}{2}v_{\mathrm{F}}^2 \frac{(\bar{\ga}+\Gph\bw^2)(\bar{\ga}\Ge+(\bar{\ga}+\Ge)\Gph\bw^2)}{\bar{\ga}^2(\Ge^2+\o^2)+2\bar{\ga}\Gph(\Ge(\bar{\ga}+\Ge)+\o^2)\bw^2+\Gph^2((\bar{\ga}+\Ge)^2+\o^2)\bw^4},\\
\sigma_{yx}&\approx e^2\frac{\nu}{2}v_{\mathrm{F}}^2 \frac{\o(\bar{\ga}+\Gph\bw^2)^2}{\bar{\ga}^2(\Ge^2+\o^2)+2\bar{\ga}\Gph(\Ge(\bar{\ga}+\Ge)+\o^2)\bw^2+\Gph^2((\bar{\ga}+\Ge)^2+\o^2)\bw^4},\\
T\bar{\kappa}_{xx}&\approx \frac{\nu}{2}p_{\mathrm{F}}^2v_{\mathrm{F}}^4 \left( \frac{\ba^2(\Ge+\bar{\ga})}{\Ge^2+\o^2+2\Ge\bar{\ga}+\bar{\ga}^2}+ \frac{r^4\bw^2( \bw^2\bar{\ga}[\Ge(\bar{\ga}+\Ge)+\o^2]+\bw^4\Gph[(\bar{\ga}+\Ge)^2+\o^2] )}{\bar{\ga}^2(\Ge^2+\o^2)+2\bar{\ga}\Gph(\Ge(\bar{\ga}+\Ge)+\o^2)\bw^2+\Gph^2((\bar{\ga}+\Ge)^2+\o^2)\bw^4} \right),\\
T\bar{\kappa}_{yx}&\approx  \frac{\nu}{2}p_{\mathrm{F}}^2v_{\mathrm{F}}^4 \left(\frac{\o\ba^2}{\Ge^2+\o^2+2\Ge\bar{\ga} +\bar{\ga}^2} + \frac{r^4\bw^4\bar{\ga}^2\o}{\bar{\ga}^2(\Ge^2+\o^2)+2\bar{\ga}\Gph(\Ge(\bar{\ga}+\Ge)+\o^2)\bw^2+\Gph^2((\bar{\ga}+\Ge)^2+\o^2)\bw^4} \right),\\
T\a_{xx}&\approx -e\frac{\nu}{2}p_{\mathrm{F}}v_{\mathrm{F}}^3 \Bigg(\frac{\bar{\gamma} r^2\bw^2(\bar{\gamma}\Ge+(\bar{\gamma}+\Ge)\Gph\bw^2)}{\bar{\gamma}^2(\Ge^2+\o^2)+2\bar{\gamma}\Gph(\Ge(\bar{\gamma}+\Ge)+\o^2)\bw^2+\Gph^2((\bar{\gamma}+\Ge)^2+\o^2)\bw^4}+ \frac{\ba^2(\Ge+\bar{\gamma})}{\Ge^2+\o^2+2\Ge\bar{\gamma} +\bar{\gamma}^2}\\
&  +\frac{\ba}{\left((\bar{\gamma}_{\e}+\bar{\gamma} )^2+\omega_c^2\right)\left((\bar{\gamma}\Gamma_{\e}+\Gamma_{\ph}(\bar{\gamma}+\Gamma_{\e})\bw^2)^2+\omega_c^2(\bar{\gamma}+\Gamma_{\ph}\bw^2)^2 \right)}\nonumber\\
&\times \Bigg\{(\bar{\gamma}+\Gamma_{\ph}\bw^2)\left[-c(2\bar{\gamma}\Gamma_{\e}\omega_c+\Gamma_{\ph}(\bar{\gamma}+2\Gamma_{\e})\omega_c\bw^2)+b(\bar{\gamma}(-\Gamma_{\e}^2+\omega_c^2)+\Gamma_{\ph}(-\Gamma_{\e}(\bar{\gamma}+\Gamma_{\e})+\omega_c^2)\bw^2)\right] \nonumber\\
& -\bar{\gamma} \left[ c\omega_c(\bar{\gamma}+\Gamma_{\ph})^2+b(\bar{\gamma}^2\Gamma_{\e}+\bar{\gamma}\Gamma_{\ph}(\bar{\gamma}+2\Gamma_{\e})\bw^2+\Gamma_{\ph}^2(\bar{\gamma}+\Gamma_{\e})\bw^4 )    \right] \nonumber\\
& -\ba\bar{\gamma}  (2\bar{\gamma}+\Gamma_{\ph}\bw^2)\left[\bar{\gamma}(-\Gamma_{\e}^2+\omega_c^2)+\Gamma_{\ph}(-\Gamma_{\e}(\bar{\gamma}+\Gamma_{\e})+\omega_c^2)\bw^2  \right] \nonumber\\
&+\ba\bar{\gamma}^2 (2\bar{\gamma}+\Gamma_{\ph}\bw^2)\left[ \Gamma_{\e}\Gamma_{\ph}\bw^2+\bar{\gamma}(\Gamma_{\e}+\Gamma_{\ph}\bw^2)\right]\Bigg\}\Bigg)\nonumber ,\\
T\a_{yx}&\approx -e\frac{\nu}{2}p_{\mathrm{F}}v_{\mathrm{F}}^3 \Bigg( \frac{\o\bar{\gamma} r^2\bw^2(\bar{\gamma}+\Gph\bw^2)}{\bar{\gamma}^2(\Ge^2+\o^2)+2\bar{\gamma}\Gph(\Ge(\bar{\gamma}+\Ge)+\o^2)\bw^2+\Gph^2((\bar{\gamma}+\Ge)^2+\o^2)\bw^4}+ \frac{\o\ba^2}{(\Ge^2+\o^2)+2\Ge\bar{\gamma} +\bar{\gamma}^2}\\
&+\frac{\ba}{\left((\Gamma_{\e}+\bar{\gamma} )^2+\omega_c^2\right)\left((\bar{\gamma}\Gamma_{\e}+\Gamma_{\ph}(\bar{\gamma}+\Gamma_{\e})\bw^2)^2+\omega_c^2(\bar{\gamma}+\Gamma_{\ph}\bw^2)^2 \right)}\nonumber \\
&\times \Bigg\{(\bar{\gamma}+\Gamma_{\ph}\bw^2)\left[-b(2\bar{\gamma}\Gamma_{\e}\omega_c+\Gamma_{\ph}(\bar{\gamma}+2\Gamma_{\e})\omega_c\bw^2)-c(\bar{\gamma}(-\Gamma_{\e}^2+\omega_c^2)+\Gamma_{\ph}(-\Gamma_{\e}(\bar{\gamma}+\Gamma_{\e})+\omega_c^2)\bw^2) \right] \nonumber \\
&- \bar{\gamma} \left[ b\omega_c(\bar{\gamma}+\Gamma_{\ph})^2-c(\bar{\gamma}^2\Gamma_{\e}+\bar{\gamma}\Gamma_{\ph}(\bar{\gamma}+2\Gamma_{\e})\bw^2+\Gamma_{\ph}^2(\bar{\gamma}+\Gamma_{\e})\bw^4 )    \right] \nonumber\\
&+ \ba\bar{\gamma} \omega_c (2\bar{\gamma}+\Gamma_{\e}\bw^2)(2\bar{\gamma}\Gamma_{\e}+(\bar{\gamma}+2\Gamma_{\e})\Gamma_{\ph}\bw^2 )\nonumber\\
&+\ba\bar{\gamma}^2 \omega_c (2\bar{\gamma}+\Gamma_{\ph}\bw^2)(\bar{\gamma}+\Gamma_{\ph}\bw^2)    \Bigg\}    \Bigg) \nonumber.
\end{align}
\end{subequations}
Notice that we still keep the leading order in $T/T_{\mathrm{F}}$. All the conductivities above can be checked to reduce to \eqnref{eqn:conductivity} when $B=0$. The experimental thermal conductivity is given by
\begin{subequations}
\begin{align}
 T\kappa_{xx}&\approx \left\{ \begin{array}{ll}
\displaystyle \frac{\nu}{2}p_{\mathrm{F}}^2v_{\mathrm{F}}^4 \left(  \frac{(\ba^2\Ge+r^2\ga)}{\Ge^2+\o^2+2\Ge\ga r^2/\ba^2 +\ga^2r^4/\ba^4}  + \frac{r^4\bw^2}{\Gamma_{\ph}+\gamma/\bw^2}  \right)  & ~ T<T_{\text{BG}}\\
\displaystyle \frac{\nu}{2}p_{\mathrm{F}}^2v_{\mathrm{F}}^4  \left( \frac{\ba^2(\Ge+\bar{\ga})}{\Ge^2+\o^2+2\Ge\bar{\ga}+\bar{\ga}^2} + \frac{r^4\bw^2}{\Gamma_{\ph}+\bar{\gamma}/\bw^2}  \right)& ~T_{\text{BG}}<T\ll T_{\mathrm{F}}\\
 \end{array} \right. ,\\
T\kappa_{yx}&\approx \left\{ \begin{array}{ll}
\displaystyle  \frac{\nu}{2}p_{\mathrm{F}}^2v_{\mathrm{F}}^4 \frac{\o\ba^2}{(\Ge^2+\o^2)+2\Ge\ga r^2/\ba^2 +\ga^2r^4/\ba^4}   & ~ T<T_{\text{BG}}\\
\displaystyle  \frac{\nu}{2}p_{\mathrm{F}}^2v_{\mathrm{F}}^4 \frac{\o\ba^2}{\Ge^2+\o^2+2\Ge\bar{\ga} +\bar{\ga}^2} & ~T_{\text{BG}}<T\ll T_{\mathrm{F}}\\
 \end{array} \right. .
\end{align}
\end{subequations}
We find that the phonon modes contribute to the dissipative experimental thermal conductivity in a  $B$-independent way, while do not contribute to the experimental thermal Hall conductivity.

\bibliography{e_ph}

\end{document}